\documentclass[fleqn,usenatbib]{mnras}

\usepackage{newtxtext,newtxmath}

\usepackage[T1]{fontenc}
\usepackage{ae,aecompl}

\usepackage{graphicx}	
\usepackage{amsmath}	
\usepackage{mathtools}
\usepackage{hyperref}
\usepackage{xspace}	
\usepackage{xcolor}	
\usepackage{euscript}
\usepackage{pdflscape}
\usepackage{afterpage}
\usepackage{placeins}
\usepackage{bm}
\usepackage{ dsfont }
\usepackage{listings}
\usepackage{soul}

\renewcommand{\t}{\mathrm{T}}
\DeclareMathOperator*{\var}{Var}

\definecolor{codegreen}{rgb}{0,0.6,0}
\definecolor{codegray}{rgb}{0.5,0.5,0.5}
\definecolor{codepurple}{rgb}{0.58,0,0.82}
\definecolor{backcolour}{rgb}{0.95,0.95,0.92}

\lstdefinestyle{python}{
    backgroundcolor=\color{backcolour},   
    commentstyle=\color{codegreen},
    keywordstyle=\color{magenta},
    numberstyle=\tiny\color{codegray},
    stringstyle=\color{codepurple},
    basicstyle=\ttfamily\footnotesize,
    breakatwhitespace=false,         
    breaklines=true,                 
    captionpos=b,                    
    keepspaces=true,                 
    numbers=left,                    
    numbersep=5pt,                  
    showspaces=false,                
    showstringspaces=false,
    showtabs=false,                  
    tabsize=2,
    upquote=true
}

\newcommand{\gaia}{{\it Gaia}\xspace}
\defcitealias{PaperI}{Paper~I}
\defcitealias{PaperII}{Paper~II}

\title[Using hidden states to gaps and the scanning law]{Completeness of the \textit{Gaia}-verse III: using hidden states to infer gaps, detection efficiencies and the scanning law from the DR2 light curves}

\author[D. Boubert, A. Everall, J. Fraser, A. Gration and B. Holl]{ 
	Douglas Boubert$^{1,2}$\thanks{E-mail: douglas.boubert@magd.ox.ac.uk},
	Andrew Everall$^{3}$,
	Jack Fraser$^{2}$,
	Amery Gration$^{2}$,
	and Berry Holl$^{4,5}$
	\\
	$^{1}$Magdalen College, University of Oxford, High Street, Oxford OX1 4AU, UK\\
	$^{2}$Rudolf Peierls Centre for Theoretical Physics, Clarendon Laboratory, Parks Road, Oxford OX1 3PU, UK\\
	$^{3}$Institute of Astronomy, University of Cambridge, Madingley Road, Cambridge CB3 0HA, UK\\
	$^{4}$ Department of Astronomy, University of Geneva, Ch. des Maillettes 51, CH-1290 Versoix, Switzerland \\
    $^{5}$ Department of Astronomy, University of Geneva, Ch. d’Ecogia 16, CH-1290 Versoix, Switzerland \\
}

\date{Accepted XXX. Received YYY; in original form ZZZ}

\pubyear{}

\begin{document}
\label{firstpage}
\pagerange{\pageref{firstpage}--\pageref{lastpage}}
\maketitle

\begin{abstract}
	The completeness of the \gaia catalogues heavily depends on the status of that space telescope through time. Stars are only published with each of the astrometric, photometric and spectroscopic data products if they are detected a minimum number of times. If there is a gap in scientific operations, a drop in the detection efficiency or \gaia deviates from the commanded scanning law, then stars will miss out on potential detections and thus be less likely to make it into the \gaia catalogues. We lay the groundwork to retrospectively ascertain the status of \gaia throughout the mission from the tens of individual measurements of the billions of stars, by developing novel methodologies to infer both the orientation and angular velocity of \gaia through time and gaps and efficiency drops in the detections. We have applied these methodologies to the \gaia DR2 variable star epoch photometry -- which are the only publicly available \gaia time-series at the present time -- and make the results publicly available. We accompany these results with a new Python package \textsc{scanninglaw} (\url{https://github.com/gaiaverse/scanninglaw}) that you can use to easily predict \gaia observation times and detection probabilities for arbitrary locations on the sky.
\end{abstract}

\begin{keywords}
	stars: statistics, Galaxy: kinematics and dynamics, Galaxy: stellar content, methods: data analysis, methods: statistical
\end{keywords}



\section{Introduction}

\gaia \citep{Gaia2016} is the first all-sky astrometric, photometric and spectroscopic survey with tens to hundreds of repeat observations over a decade-long baseline. Future \gaia data releases will drive a generational breakthrough in our ability to carry out time-domain astrophysics, allowing us to study binary stars, exoplanets, variable stars, supernovae and microlensing events through combinations of epoch astrometric, photometric and spectroscopic measurements. Our ability to discover a transient source or detect the variability of a periodic source will depend on both the number of \gaia observations and the timings between them (as well as the uncertainties of the individual observations). The pattern of these \gaia observations varies systematically across the sky -- as described by the scanning law which states where \gaia was looking at every point in time -- and thus we will be differently able to investigate time-domain phenomena in different regions of the sky. The picture is further complicated by the possibility that a \gaia observation predicted by the scanning law may not result in a published measurement, as that can change the completeness and sensitivity of the \gaia pipeline to time variable objects at that location on the sky. The set of variable and non-variable objects in the \gaia catalogues will be biased by the selection function resulting from the effective scanning law that was used, and we will not be able to make unbiased general statements unless we can correct for these biases, which will require us (amongst other things) to know when \gaia looked at a given point on the sky and the probability that each of those observations would have resulted in a measurement of a given star. It is thus of fundamental importance that we have a precise knowledge of the \gaia scanning law, the reasons why a \gaia observation might not result in a published measurement, and the times when no \gaia measurements were acquired.

In \citet{PaperI} (hereafter referred to as \citetalias{PaperI}) we conducted preliminary work towards that goal by exploiting the light curves of the 550,737 variable sources published in \gaia DR2 \citep{Gaia2018, Holl2018,Riello2018, Evans2018}, which tell us the time of each detection and the location of the detection on the \gaia focal plane. By aligning the reported location of the source on the focal plane with the location of the source on the sky, we were able to derive first-order corrections to the \gaia   ~DR2 nominal scanning law\footnote{\url{https://www.cosmos.esa.int/web/gaia/scanning-law-pointings}} (the scanning law that DPAC commanded \gaia to follow and which can deviate from the true scanning law by up to $30\;\mathrm{arcsec}$). By looking at the times between consecutive photometric measurements of all of the variable stars, and by comparing the predicted number of observations at each location on the sky to the reported maximum number of astrometric, photometric and spectroscopic detections, we were able to identify `gaps' which contributed no detections to the published catalogue.

There were several deficiencies in that work which we will address here. First, our corrections to the scanning law were simplistic and only corrected the pitch and roll of \gaia, not the yaw. Our corrected scanning law was thus able to predict the across-scan location of the source to a precision of $0.083\;\mathrm{arcsec}$, but the along-scan location remained uncertain at a level of $30\;\mathrm{arcsec}$, with a corresponding uncertainty on the timing of predicted observations of $0.5\;\mathrm{sec}$. Second, we defined a gap to have occurred between any two consecutive photometric measurements more than 1\% of a day apart without a robust statistical justification. This may have caused us to both miss shorter gaps and to erroneously interpret long intervals between consecutive observations as gaps in regions of the sky that are simply sparse in variable stars. Third, we ignored that another reason for a \gaia observation not to have resulted in a published detection is if that detection is deleted on-board due to \gaia scanning 
dense regions of the sky for large fractions of the spin period and exhausting its available storage capacity due to the limited down-link bandwidth. This will result in a magnitude dependent  deletion fraction that varies as a function of time as \gaia scans across more and less dense parts of the sky. Fourth, we ignored that there are other processes besides gaps and on-board deletions that can cause an observation to not result in a published detection, ranging from faint sources not being detected every time due to photon shot noise to crowding preventing every source from being assigned a window to quality cuts applied by the \gaia Data Processing and Analysis Consortium (DPAC) as listed in the Introduction of \citetalias{PaperI}. 

It is likely that we will not be able to identify with full certainty the nature of all of the processes that caused observations to be missed. Instead, our goal in this work is to characterise the time-varying and partially-probabilistic process through which \gaia observations result in published detections, to the best degree possible with the 550,737 light-curves associated with \gaia DR2 variable sources. Our methods will be immediately applicable to larger samples of \gaia epoch measurements when they are made available in later data releases. The fourth and subsequent \gaia data releases will contain all of the epoch astrometric, photometric and spectroscopic measurements and an extension of this methodology will allow the community to diagnose the photometric and temporal biases impairing the completeness of these data releases.

This work consists of two parts. In Sec. \ref{sec:scanninglaw} we use the locations of the variable stars both on the focal plane and on the sky at their times of detection to infer \gaia's orientation and angular velocity throughout the twenty two months of \gaia DR2. In Sec. \ref{sec:fractions} we combine the detection times of the variable stars with the predicted observations that did not result in published detections to simultaneously infer gaps which did not result in any published detections and the magnitude-dependency of the detection probability with time. The methodologies in both sections are based on temporally-evolving state space models and so we begin with a pedagogical overview of Gaussian process regression, Hidden Markov Models and Kalman Filters in Sec. \ref{sec:pedagogy}. 

\section{Modelling states through time}
\label{sec:pedagogy}

Throughout this paper we will consider probabilistic models with time-evolving state parameters $\vec{\theta}(t)$ that describe some aspect of a process that we are interested in. At discrete times $\{t_{i}\}_{i = 1}^{n}$ we make a noisy measurement of either those parameters, a function of those parameters, or random variables that are conditioned on those parameters. The latter two cases are known as hidden state spaces, because we never have direct access to the values of the state parameters. In this section we give an overview of such models as a precursor to the other sections in this work.

\subsection{Gaussian process regression}
\label{sec:gaussianprocesses}


Perhaps the most common state-space probabilistic model is the Gaussian process regression model, which allows us to predict a state at some arbitrary time given observations of that state at some number of other times \citep{sacks1989}. 
We treat each state $\theta(t)$ as the realization of a random variable $\Theta(t)$ and assume that for any finite set of times $\{t_i\}_{i=1}^{n}$ those random variables $\{\Theta(t_{i})\}_{i=1}^{n}$ have a multivariate normal distribution.
Thus, the time evolution of a state, $\{\theta(t)\}_{t \in \mathbb{R}}$, is a realization of a Gaussian process.
A Gaussian random process is completely defined by its mean-value function, $m$, given by $m(t) = \langle\Theta(t)\rangle$, and its covariance kernel, $k$, given by $k(t, t') = \var(\Theta(t), \Theta(t'))$.
Accordingly, we write $\Theta \sim \operatorname{GP}(m, k)$.
Given a sample of $\{\Theta(t_{i})\}_{i = 1}^{n}$, which we can express as a real vector $\vec{\theta} = [\theta(t_{1}) \dots \theta(t_{n})]^{\t}$, we may predict the random variable $\Theta(t_{*})$, for any arbitrary time $t_{*}$, using the conditional random variable $\Theta(t_{*})\;|\;(\{\Theta(t_{i})\}_{i = 1}^{n} = \vec{\theta})$.
We denote this random variable $\hat{\Theta}(t_{*})$.
It may be shown \citep{rasmussen2006} that $\hat{\Theta}(t_{*})$ is normal with mean
\begin{align}
  \langle\hat{\Theta}(t_{*})\rangle = m(t_{*}) + \vec{k}^{\t}\vec{K}^{-1}(\vec{\theta} - \vec{m})
\end{align}
and variance
\begin{align}
    \var(\hat{\Theta}(t_{*})) = k(t_{*}, t_{*}) - \vec{k}^{\t}K^{-1}\vec{k}
\end{align}
where $\vec{k} = [k(t_{*}, t_{1}) \dots k(t_{*}, t_{n})]^{\t}$,
$K_{ij} = k(t_{i}, t_{j})$, and
$\vec{m} = [m(t_{1}) \dots m(t_{n})]^{\t}$.
Typically, we know neither the mean-value function nor the covariance kernel.
Instead, we assume that they are members of some family of functions, which is parameterised by a set of hyperparameters.

It is common to assume that the mean is constant, given by $m(t) = \mu$ for some real $\mu$,
and that the covariance kernel is a linear sum of a squared exponential and a Kronecker delta, given by
\begin{align}
    \label{eq:gp_cov_kernel}
    k(t, t') = \varepsilon^{2}\exp\left(\dfrac{-(t' - t)^{2}}{2l^{2}}\right) + \sigma^2(t)\delta(t, t').
\end{align}

The first term of this expression is the covariance kernel for the underlying time-series, and the second is the covariance kernel for the noise in each observation, with $\sigma(t_i)$ being the known noise of the measurement at time $t_i$. We call the hyperparameter $\mu$ the `signal mean', $\varepsilon^{2}$ the `signal variance' and $l$ the `time scale', and collate them in the hyperparameter vector $\vec{\xi}=[\mu,\varepsilon^2,l]$.

By maximising the likelihood we can recover the hyperparameters that best suits the data. The random variables $\{\Theta(t_{i})\}_{i = 1}^{n}$ have a multivariate normal distribution, and therefore have the probability density function (PDF)
\begin{align}
    f(\vec{\theta}; \vec{\xi})
    &= \dfrac{1}{\sqrt{{(2\pi)^{n}|K_{\vec{\xi}}|}}} \exp\left(-\dfrac{1}{2}\left(\vec{\theta} - \vec{m}_{\vec{\xi}}\right)^{\t}{K_{\vec{\xi}}}^{\,-1}\left(\vec{\theta} - \vec{m}_{\vec{\xi}}\right)\right),
\end{align}
where the subscripts now indicate that $K$ and $\vec{m}$ are functions of the hyperparameter vector $\vec{\xi}$.
By definition, the likelihood of $\vec{\xi}$ is given by
\begin{align}
    L\left(\vec{\xi}| \vec{\theta}\right) = f\left(\vec{\theta}| \vec{\xi}\right)
\end{align}
and hence the log-likelihood of $\vec{\xi}$ is, up to an additive  constant,
\begin{align}
    \ln\left(L\left(\vec{\xi}| \vec{\theta}\right)\right)
    &= - \dfrac{1}{2} \left(\vec{\theta} - \vec{m}_{\vec{\xi}}\right)^{\t}{K_{\vec{\xi}}}^{\,-1}\left(\vec{\theta} - \vec{m}_{\vec{\xi}}\right) - \dfrac{1}{2}\ln\left(|K_{\vec{\xi}}|\right). 
    \label{eq:lnL_prediction}
\end{align}
We must therefore find the hyperparameter vector that maximizes this expression.
It is common practice to ignore the error associated with the resulting maximum-likelihood estimate of the hyperparameter vector, and to proceed as if the hyperparameter is known with certainty.

When the mean is constant we may simply estimate it by setting it equal to the sample mean or -- more accurately -- the weighted sample mean, 
\begin{align}
    \label{eq:gp_sample_mean}
    \mu = \dfrac{\sum_{i = 1}^{n}\theta_{i}/\sigma_i}{\sum_{i = 1}^{n}1/\sigma_{i}}
\end{align}
where $\theta_i$ and $\sigma_i$ are the reported measurement and uncertainty at time $t_i$.
When the signal variance dominates the noise variance, i.e.\ when $\varepsilon^{2} \gg \sigma^{2}$, we may estimate it by setting it equal to the sample variance or -- more accurately -- the weighted sample variance,
\begin{align}
\label{eq:gp_sample_variance}
    \varepsilon^{2} &= \dfrac{\sum_{i = 1}^{n}(\theta_{i} - \mu)^{2}/\sigma_{i}}{\sum_{i = 1}^{n}1/\sigma_{i}}.
\end{align}
Similarly, the time scale, $l$, can be set equal to some known characteristic time scale for the physical process in question.
Again, it is common practice to ignore the error associated with these estimates.

We will use Gaussian process regression in Sec. \ref{sec:predictingobs} to fill in missing values in the \gaia DR2 variable star light-curves, by inferring the magnitudes of the stars at the predicted observation times that did not yield published detections.

\subsection{Hidden Markov Models}
\label{sec:hiddenmarkov}

Hidden Markov Models are used to model a system that is known to transition between discrete states through a Markov Process $X$, but where we do not have direct access to the value of the states $\{X_i\}_{i=1}^n$. There is assumed to be another discrete process $Y$ whose states $\{Y_i\}_{i=1}^n$ are conditioned on process $X$ and are observed, and so the task is to infer the states of process $X$ given the observed states of process $Y$. For instance, the lead author regularly travels by train from Oxford to London. Whether that train leaves Oxford on time on a particular day ($Y_i=1$) or not ($Y_i=0$) is conditionally dependent on whether there were leaves on the track between Hereford and Oxford ($X_i=1$) or not ($X_i=0$). The lead author will not know if there were leaves on the track, but the train is more likely to be running late if this was the case. The probability of there being leaves on the track will depend in some way on whether there were leaves on the track the previous day, for instance through the weather or the number of leaves left on the trees.

The Markov property tells us that the probability distribution of the future state $X_{i}$ conditioned on the values of all previous states only depends on the present state,
\begin{equation}
    P(X_{i}|X_1=x_1,\dots,X_{i-1}=x_{i-1}) = P(X_{i}|X_{i-1}=x_{i-1}).
\end{equation}
The other key property of a Hidden Markov Process is that the observation at the present time $Y_i$ is only conditionally dependent on the present hidden state,
\begin{equation}
    P(Y_{i}|X_1=x_1,\dots,X_i=x_i) = P(Y_{i}|X_i=x_i).
\end{equation}
These two probability distributions defined at each step completely specify a Hidden Markov Model. In the analogy above, we would need to specify the probability of there being leaves on the track today given whether there were leaves on the track yesterday, and the probability of the Oxford-to-London train leaving Oxford on time depending on whether there are leaves on the track today.

It is common for these probability distributions to be parameterised by hyper-parameters $\vec{\xi}$, whose values are optimised by maximising the likelihood of the observations. By marginalising over the hidden states it is possible to condition the observation at the next state on the observations at all previous states $P(Y_{i}|Y_1=y_1,\dots,Y_{i-1}=y_{i-1},\vec{\xi})$ and thus compute the likelihood of the hyper-parameters $\vec{\xi}$ as 
\begin{align}
    L(\vec{\xi}|Y_1&=y_1,\dots,Y_n=y_n) \nonumber \\
    &= P(Y_1=y_1|\vec{\xi})\prod_{i=2}^n P(Y_i|Y_1=y_1,\dots,Y_{i-1}=y_{i-1},\vec{\xi}).
\end{align}
The Viterbi algorithm \citep{Viterbi1967} can be used to find the most likely sequence of hidden state values given the observations and fixed values for any hyper-parameters. We stress that there is a difference between the most likely value of a state at each time and the value of the state in the most likely sequence of states.

We will use a Hidden Markov Model in Sec. \ref{sec:discrete} to identify gaps when \gaia observations were not resulting in published detections.

\subsection{Kalman filters}
\label{sec:kalmanfilters}

Kalman filters are an algorithm to estimate a hidden dynamical state $\vec{x}(t)$ at a series of discrete times ${t_{i=1,\dots,n}}$ under two assumptions. First, the state at time $t_{i}$ is a linear function of the state at $t_{i-1}$,
\begin{equation}
    \vec{x}_{i} = F_i\vec{x}_{i-1} + \vec{w}_{i},
\end{equation}
where $F_i$ is the matrix describing the state-transition model and $\vec{w}_{i}$ is the process noise which is assumed to be drawn from a zero-mean multivariate normal distribution with covariance matrix $Q_k$. Second, at each time a noisy measurement $\vec{z}_i$ is made of the true state $\vec{x}_i$ through the linear model
\begin{equation}
    \vec{z}_{i} = H_i\vec{x}_{i} + \vec{v}_{i},
\end{equation}
where $H_i$ is the matrix describing the observation model and $\vec{v}_{i}$ is the observation noise which is assumed to be drawn from a zero-mean multivariate normal distribution with covariance matrix $R_k$.

The Kalman filter represents our estimate of the state through a multivariate normal distribution with mean $\vec{\hat{x}}_{i|j}$ and covariance $P_{i|j}$, which are the mean and covariance of our estimate at time $t_i$ given all of the measurements made at times up to and including time $t_j$. The Kalman filter iteratively calculates the means $\vec{\hat{x}}_{i|i}$ and covariances $P_{i|i}$ at each time-step given all of the prior observations. Each time-step is split into a prediction stage (where $\vec{\hat{x}}_{i|i-1}$ and $P_{i|i-1}$ are predicted from $\vec{\hat{x}}_{i-1|i-1}$ and $P_{i-1|i-1}$) and a measurement update stage (where the updated estimates $\vec{\hat{x}}_{i|i}$ and $P_{i|i}$ are obtained based on the measurement $\vec{z}_i$). The full system of equations is
\begin{align}
    \vec{\hat{x}}_{i|i-1} &= F_k\vec{\hat{x}}_{i-1|i-1} \\
    P_{i|i-1} &= F_iP_{i-1|i-1}F_i^T+Q_i \\
    \vec{y}_i &\equiv \vec{z}_i - H_i\vec{\hat{x}}_{i|i-1} \\
    S_i &\equiv H_iP_{i|i-1}H_i^T + R_i \\
    K_i &\equiv P_{i|i-1}H_i^TS_i^{-1} \\
    \vec{\hat{x}}_{i|i} &= \vec{\hat{x}}_{i|i-1} + K_i \vec{y}_i \\
    P_{i|i} &= (I-K_iH_i)P_{i|i-1}.
\end{align}
The terms $\vec{y}_i$ and $S_i$ are the mean and covariance of the residual between the measurement and the prediction, while the term $K_i$ is the matrix of optimal weights to apply to the residual when computing the updated estimate, otherwise known as the Kalman gain.
If the period under consideration is a fixed-interval $t_1\leq t \leq t_n$, then it is possible to obtain an improved estimate of the state at time $t_i$ by incorporating the information from observations at later times $t_i<t_j\leq t_n$. This is done through the Rauch-Tung-Striebel smoothing algorithm \citep{Rauch1965} which performs a backward pass starting at time $t_n$, where we note that the final mean and covariance from the forward algorithm $\vec{\hat{x}}_{n|n}$ and $P_{n|n}$ already utilise all of the available information. At time $i$ the system of equations to be evaluated is
\begin{align}
    C_i &= P_{i|i}F_{i+1}^TP_{i+1|i}^{-1} \\
    \vec{\hat{x}}_{i|n} &= \vec{\hat{x}}_{i|i} + C_i\left(\vec{\hat{x}}_{i+1|n} -  \vec{\hat{x}}_{i+1|i}\right) \\
    P_{i|n} &= P_{i|i} + C_i \left(P_{i+1|n} - P_{i+1|i}\right)C_i^T.
\end{align}
A toy example of a system where Kalman filters are suitable is in estimating the position and velocity of a moving object subject to random, zero-mean accelerations where we only have access to occasional, noisy measurements of the position. The state encodes the position $p_i$  and velocity $q_i$ at each time $t_i$, and the expected position and velocity at the next time is a linear function of the estimate of the position and velocity at the previous time
\begin{align}
    \hat{p}_{i+1} &= \hat{p}_i + \hat{q}_i\Delta t_i, \\
    \hat{q}_{i+1} &= \hat{q}_i,
\end{align}
where $\Delta t_i = t_i-t_{i-1}$.
In the formalism above, we would write
\begin{equation}
    \vec{x}_i=\begin{bmatrix}p_i \\ q_i\end{bmatrix},\quad F_i = \begin{bmatrix} 1 & \Delta t_i \\ 0& 1\end{bmatrix}, \quad H_i = \begin{bmatrix} 1 & 0 \end{bmatrix}.
\end{equation}

The Kalman filter can be generalised to non-linear dynamical systems
\begin{align}
    \vec{x}_{i} &= f(\vec{x}_{i-1}) + \vec{w}_{i} \\
    \vec{z}_{i} &= h(\vec{x}_{i}) + \vec{v}_{i},
\end{align}
in which case it is known as the extended Kalman filter. The extended Kalman filter algorithm only differs from the linear case in that the prediction of the next state and the expected observation use the non-linear functions
\begin{align}
    \vec{\hat{x}}_{i|i-1} &= f(\vec{\hat{x}}_{i-1|i-1}) \\
    \vec{y}_i &= \vec{z}_i - h(\vec{\hat{x}}_{i|i-1}),
\end{align}
whilst the remaining equations remain the same with the matrices $F_i$ and $H_i$ being defined by the Jacobians
\begin{equation}
    F_i = \frac{\partial f}{\partial \vec{x}}\bigg\rvert_{\vec{\hat{x}}_{i-1|i-1}}, \quad H_i = \frac{\partial h}{\partial \vec{x}}\bigg\rvert_{\vec{\hat{x}}_{i|i-1}}.
\end{equation}
A modification of the extended Kalman filter to allow for multiplicative uncertainties will be presented in Sec. \ref{sec:mekf} to allow us to model the orientation of \gaia, whilst an extended Kalman filter will be used in Sec. \ref{sec:continuous} to model the efficiency with which \gaia observations result in detections through time.

\section{Scanning law}
\label{sec:scanninglaw}

Determining the \gaia scanning law is exactly analogous to the problem of determining the attitude of \gaia through time. Determining the attitude of a spacecraft is a key engineering challenge which has been covered extensively in the literature (see \citealp{Wertz2012} and references therein), although our use case is peculiar in that we do not have access to gyroscopic\footnote{Note that \gaia does not use any gyroscopes to guide its attitude, instead it uses a cold-gas micro-newton thruster system firing several times per second, using the scientific instrument measurements to maintain the programmed scanning law during nominal operations \citep{Gaia2016}.} measurements or data from a star-tracker. Instead, we will attempting to infer the attitude of \gaia based on the nominal scanning law and the locations and timings of the detections of the \gaia DR2 variable stars. Note that in the post-processed Astrometric Global Iterative Solution of \gaia \cite[AGIS, see][]{Lindegren:2012aa}  the attitude is also derived from the observations themselves, though based on a set of `primary' sources that behave like single point-like sources whose colour is stable. \gaia's point spread function is colour-dependent \citep{Lindegren2018} and so if a source's colour changes with time then that adds additional uncertainty into the determination of its position.
Variable stars have colour variations and could be components of astrometric binary systems, both of which make them less than ideal as reference sources for the extreme attitude precision that the official astrometric solution requires, but are the only sources for which we have \gaia time-series. We chose to use the first-order Multiplicative Extended Kalman Filter \citep{Markley2003} where the attitude of the spacecraft is expressed as a quaternion, the attitude error is multiplicative, and the attitude error and angular velocity vector are assumed to be hidden states that jointly satisfy a multivariate normal distribution at all times. We use quaternions as they are the ideal mathematical object to encode the orientation of a spacecraft (an argument developed in the following section) and were used in the AGIS \citep{Lindegren:2012aa} to determine Gaia's attitude for the official \gaia DPAC astrometric solution.

\subsection{What are quaternions?}
Quaternions ($z=a+bi+cj+dk$) are a generalisation of the complex numbers ($z=a+bi$) with two additional `imaginary' dimensions satisfying $i^2=j^2=k^2=ijk=-1$. If this definition is as impenetrable to the reader as it was to the authors, then we recommend an hour exploring \citet{3blue1brown}. An alternative way to write quaternions is in the composite form $\mathbf{q}=(\vec{q},q_4)$, where $q_4 \equiv a$ is the real part of the quaternion, and $\vec{q} \equiv (b,c,d)^T$ corresponds to the imaginary parts. The key utility of quaternions is that any rotation by an angle $\theta$ about an axis specified by the unit vector $\hat{u}$ in three real dimensions can be uniquely encoded, without the risk of gimbal lock, as a single number in quaternion space
\begin{equation}
    \mathbf{q} = \cos{\frac{\theta}{2}} + \left(u_xi+u_yj+u_zk \right)\sin{\frac{\theta}{2}}.\label{eq:RotationQuaternions}
\end{equation}
Every unit quaternion (e.g. a quaternion normalised such that $|\vec{q}|^2+q_4^2=1$) corresponds to a rotation. The more common way to represent rotations in three dimensions is in terms of a rotation matrix, which can be recovered from the quaternion through
\begin{equation}
    A(\mathbf{q}) = (q_4^2-|\vec{q}|^2)I_3 + 2\vec{q}\vec{q}^T - 2q_4[\vec{q}\times],
\end{equation}
where $[\vec{q}\times]$ is the cross-product matrix
\begin{equation}
    [\vec{q}\times] \equiv 
    \begin{bmatrix}
    0 & -q_3 & q_2 \\
    q_3 & 0 & -q_1 \\
    -q_2 & q_1 & 0 \\
    \end{bmatrix}.
\end{equation}
We will follow \citet{Markley2003} in adopting a slightly altered definition of quaternion multiplication which follows that of direction cosine matrices,
\begin{align}
    \mathbf{p} &\otimes \mathbf{q} \equiv \begin{bmatrix}
    p_4\vec{q}+q_4\vec{p}-\vec{p}\times\vec{q} \\
    p_4q_4-\vec{p}\cdot\vec{q} \\
    \end{bmatrix} \\
    &\Rightarrow A(\mathbf{p})A(\mathbf{q})=A(\mathbf{p}\otimes\mathbf{q}),
\end{align}
noting that quaternion multiplication is always non-commutative.
This differs from the standard convention only up to the ordering of the quaternions: $\mathbf{qp} = \mathbf{p} \otimes \mathbf{q}$. 

The orientation of a body that is rotating can be parameterised by the required rotation from a non-rotating frame, and thus we can use a quaternion to encode the orientation of a body. Under our convention, the kinematic equation relating the time-evolution of the quaternion to the angular velocity vector $\vec{\omega}$ is
\begin{equation}
    \dot{\mathbf{q}} = \frac{1}{2} \begin{bmatrix}\vec{\omega} \\ 0\end{bmatrix} \otimes \mathbf{q}.
\end{equation}
Here $\vec{\omega}$ is the angular velocity evaluated in the frame which moves with the body such that the moment of inertia tensor is constant and diagonal. Integrating this equation of motion is non-trivial due to the manifold-constraint that the quaternion must at all times remain normalized to unity, and must therefore evolve smoothly across the unit hyper-sphere, $S^3$. Adapting the first-order manifold integrator of \citet{Crouch1993} to a quaternion framework, we find that if $\vec{\omega}$ is constant between $t$ and $t + \delta t$, then the system evolves exactly according to:
\begin{equation}
    \mathbf{q}(t+\delta t) = \begin{bmatrix} \frac{\vec{\omega}}{|\vec{\omega}|}\sin(\frac{\delta t}{2}|\vec{\omega}|) \\ \cos(\frac{\delta t}{2}|\vec{\omega}|) \end{bmatrix}\otimes\mathbf{q}(t).
\end{equation}
These expressions perfectly preserve the unit normalisation of the quaternion. In the limit of small time-steps, this expression reduces to the result of a naive additive integration of the equation of motion.

A further consequence of the unity-normalization constraint is that if the observed attitude $\mathbf{q}$ is an uncertain estimate of the true attitude $\hat{\mathbf{q}}$, then it is meaningless to interpret that uncertainty as additive $\mathbf{q} = \hat{\mathbf{q}}+\delta\mathbf{q}$, because there is no guarantee that the sum of the true attitude and noise will have unit normalisation, and hence will not represent a valid orientation state. An alternative used by \citet{Markley2003} is to consider a multiplicative uncertainty $\mathbf{q} = \boldsymbol{\delta q}(\vec{a})\otimes \hat{\mathbf{q}}$, where the uncertainty is parameterised by an unconstrained vector $\vec{a}$ in $\mathds{R}^3$ which is mapped to the space of unit quaternions by the function
\begin{equation}
    \boldsymbol{\delta q}(\vec{a}) = \frac{1}{\sqrt{4+|\vec{a}|^2}}\begin{bmatrix}\vec{a}\\2\end{bmatrix}. \label{eq:attitudeerror}
\end{equation}
The product of two unit quaternions is a unit quaternion and thus this form is guaranteed to preserve the unit normalisation. To first order in the components of $\vec{a}$, the rotation matrix corresponding to the mean rotation and an additional small rotation can be approximated by
\begin{align}
    A(\textbf{q})&=A(\boldsymbol{\delta q}(\vec{a})\otimes\hat{\textbf{q}}) \nonumber \\
    &= A(\boldsymbol{\delta q}(\vec{a}))A(\hat{\textbf{q}}) \nonumber \\
    &= (I_3-[\vec{a}\times])A(\hat{\textbf{q}}).
\end{align}

\subsection{Positions of stars on the sky and on the focal plane}
\label{sec:mekfstars}

We are attempting to infer the attitude of \gaia with very little information. Traditionally, the attitude of a spacecraft is inferred using measurements from a combination of gyroscopes and star-trackers, but we do not have access to these. We have only access to the light-curves of the 550,737 stars classified as photometrically variable by DPAC in \gaia DR2. Each of the flux measurements in those light-curves has an associated \textsc{transit}\_\textsc{id} which encodes the time $t$, the field of view $F$, and the across-scan pixel $P$ and CCD $C$ of the detection on the \gaia focal plane. In the \gaia body frame, the along-scan angles are the longitude and the across-scan angles are the latitude, so named because they point in directions parallel and perpendicular to the direction in which \gaia is scanning. Note that technically the \textsc{transit}\_\textsc{id} encodes the centre of the \textit{window} that was assigned on-board to record the flux around the identified point-source. Both time and position in the \textsc{transit}\_\textsc{id} are from the on-board estimates, which have an uncertainty of roughly 1 pixel, i.e., $\sim$60 mas in the along-scan direction (corresponding to $\sim$1~ms) and $\sim$180 mas in the across-scan direction. As we do not have access to the sub-pixel centroid estimate for each window, the precision of each transit observation will hence be limited to about 1 pixel in both directions. The timings of the detections are those measured on-board \gaia and so will be subject to aberration effects, while the timings of the brightest detections ($G<12$) could be offset due to the partial, gated integrations used to prevent saturation. 

It is more important to accurately know the across-scan location of a source than the along-scan location, because in the latter case a $20\;\mathrm{arcsec}$ offset will only cause an error of $0.3\;\mathrm{sec}$ in the observation time, while in the former case it might mean that the star falls outside of \gaia's field of view. Relativistic aberration is caused by \gaia's motion around the Sun of roughly $30\;\mathrm{km}\;\mathrm{s}^{-1}$ (\gaia is located at L2 and so moves at approximately Earth's orbital velocity), which can result in a maximum total aberration of about $20\;\mathrm{arcsec}$ \cite[see Eq.~17 of][]{klioner2003} across a combination of the along-scan and across-scan directions. Due to the importance of the across-scan location being correct, we corrected for the effect of aberration as described later in this section. The gating can only cause the detection location to be off in the along-scan direction by at most half of a CCD integration time (the ungated CCD integration time is $4.42\;\mathrm{sec}$ \cite[see Sec. 2.2 of][]{Crowley2016}) and so the predicted observation times of the gated brightest stars can be off by at most $2.2\;\mathrm{sec}$. Errors of these scales in the predicted observation time will not be astrophysically important even for stars exploding as supernovae, justifying us to neglect this effect.

Given the geometry of the \gaia focal plane, there are functions $\eta(F,P,C)$ and $\zeta(F,P,C)$ that map these quantities to along-scan and across-scan angles in the body frame of \gaia, which can then be expressed as a unit vector. The attitude of \gaia at the time of detection can then be used to rotate that unit vector to the ICRS reference frame. However, we already know the location of the star on the sky from the \gaia DR2 source catalogue and thus we can constrain the unknown attitude at the time of the detection by requiring that the location of the star on the focal plane aligns with the location of the star on the sky. We note that -- even if the position of the star on the sky and the geometry of the focal plane is known perfectly -- the alignment of one vector in the body frame with one in the reference frame only constrains two of the three degrees of freedom in the attitude.

We used the information in the \gaia DR2 light-curves in \citetalias{PaperI} to make a determination of the \gaia scanning law. We did not carry out a full attitude determination in that preliminary work, restricting ourselves to only considering across-scan corrections to the positions of the preceding and following fields-of-view. This is equivalent to changing the roll and pitch of \gaia in the body frame but leaving the yaw fixed, and thus reduces the dimensionality of the problem to two degrees of freedom. In this work we adopt the same across-scan focal plane geometry as in \citetalias{PaperI} and additionally constrain the along-scan location,
\begin{align}
    \zeta(F&,C,P) = \nonumber \\
    &\Delta_f(1-2F)+(\Delta_c+\delta_c)(C-3)-(\Delta_p+\delta_p)(P-1965/2), \nonumber \\
    \eta(F&,C,P) = \eta_{\mathrm{AF1}}+\gamma_c(1-2F) \label{eq:model}
\end{align}
where $\Delta_f=220.9979\;\mathrm{arcsec}$ is the magnitude of the across-scan offset of each of the fields-of-view, $\Delta_c=356.5435\;\mathrm{arcsec}$ is the across-scan size of each CCD,  $\Delta_p=0.1768\;\mathrm{arcsec}$ is the across-scan size of each pixel, $\eta_{\mathrm{AF1}}=-0.5^{\circ}$ is the along-scan distance between the centre of each field-of-view and the reference centre of the AF1 CCDs, and $\gamma_c = 106.5^{\circ}/2$ is half the basic angle of \gaia. The quantities $\delta_c$ and $\delta_p$ are free parameters that can correct for small errors in the across-scan sizes of the CCDs and pixels, respectively. We do not include corrections to the across-scan field-of-view offset nor the along-scan offset of AF1 from the field-of-view centres, because these are fully degenerate with the roll and yaw of \gaia. The finite size of the along-scan and across-scan pixels means that both of the expressions in Eq. \ref{eq:model} give only an estimate of the location on the focal plane. We assume a variance in the along-scan direction of $\sigma_{\eta}^2=(17\;\mathrm{arcsec})^2$ (based on the variance in the along-scan locations of stars at their times of observation if we use the nominal scanning law) and in the across-scan direction of $\sigma_{\zeta}^2 = \frac{1}{12}(\Delta_p+\delta_p)^2+\sigma_p^2$ (the first term is the variance of a uniform distribution and the second term accounts for any excess spread in the across-scan direction).

Suppose you have a mean longitude and latitude $(\hat{\phi},\hat{\theta})$ with covariant uncertainties $(\Delta\phi \cos{\hat{\theta}}, \Delta\theta)\sim N(0,\Sigma)$. To first order, the unit vector defined by those coordinates is then distributed like $\vec{u} \sim N(\hat{\vec{u}},T\Sigma T^T)$, where
\begin{equation}
    \hat{\vec{u}} = \begin{bmatrix}\cos{\hat{\phi}}\cos{\hat{\theta}} \\ \sin{\hat{\phi}}\cos{\hat{\theta}} \\ \sin{\hat{\theta}}\end{bmatrix}, \quad T = \begin{bmatrix} -\sin{\hat{\phi}} & -\cos{\hat{\phi}}\sin{\hat{\theta}} \\ +\cos{\hat{\phi}} & -\sin{\hat{\phi}}\sin{\hat{\theta}} \\ 0 & +\cos{\hat{\theta}}\end{bmatrix}.
\end{equation}
This operation is equivalent to projecting the uncertainty in the unit vector into the plane that is tangent to the unit sphere at the mean position and thus the covariance matrix $T\Sigma T^T$ is rank-deficient -- it describes uncertainty in three dimensions but only has two degrees of freedom. In almost all places where we will use this approximation the uncertainty on the positions is less than $1\;\mathrm{arcsec}$ and so the approximation is highly valid. We use these equations to express the uncertain along-scan and across-scan coordinates as a unit vector in \gaia's body frame. One of the conveniences of working with unit vectors rather than angles is that is that it is easy to rotate a mean vector $\hat{\vec{u}}$ and covariance matrix $\Sigma$ from the reference frame to the body frame of \gaia using the matrix $A(\textbf{q})$, with the rotated vector having mean $A\hat{\vec{u}}$ and covariance $A\Sigma A^T$.

The \gaia DR2 astrometry of a star is given as the mean position, parallax and proper motions $\vec{x}_0=(\hat{\alpha}\cos{\hat{\delta}},\hat{\delta},\hat{\varpi},\hat{\mu}_{\alpha}\cos{\hat{\delta}},\hat{\mu}_{\delta})_0$ at the epoch $T_0=$~J2015.5 and a joint covariance matrix $\Sigma$ describing the uncertainty in each of these. In general we will be interested in the position of the star at a different epoch $T_0+\Delta T$ and so we need to propogate the position forward or backward in time by applying the linear operator
\begin{equation}
\footnotesize
\setlength{\arraycolsep}{2.5pt}
\medmuskip = 1mu
    M = \begin{bmatrix}1 & 0 & X_{\mathrm{G}}\sin{\hat{\alpha}_0}-Y_{\mathrm{G}}\cos{\hat{\alpha}_0} & \Delta T & 0 \\
                       0 & 1 & (X_{\mathrm{G}}\cos{\hat{\alpha}_0} + Y_{\mathrm{G}}\sin{\hat{\alpha}_0})\sin{\hat{\delta}_0} - Z_{\mathrm{G}}\cos{\hat{\delta}_0} & 0 & \Delta T \\
                       0 & 0 & 1 & 0 & 0 \\
                       0 & 0 & 0 & 1 & 0 \\
                       0 & 0 & 0 & 0 & 1 
    \end{bmatrix},
\end{equation}
where $\vec{x}_{\mathrm{G}} = (X_{\mathrm{G}},Y_{\mathrm{G}},Z_{\mathrm{G}})$ is the vector between the Solar System barycentre and \gaia at the epoch $T_0+\Delta T$ (obtained from the NASA JPL HORIZONS ephemeris calculator) and we assume that the proper motions and parallax are constant over the $\pm1\;\mathrm{year}$ interval of $\Delta T$. The predicted astrometry at the epoch $T_0+\Delta T$ will have mean $M\vec{x}_0$ and covariance $M\Sigma M^T$. We predict the astrometry for all of the sources with variable light-curves in \gaia DR2 at the time of each flux measurement. We discarded the measurements from any source which lacked 5D astrometry in \gaia DR2, which resulted in us discarding 466,706 out of the 17,672,340 flux measurements. Focusing only on the rows and columns of the predicted astrometry that contain the positions, we can then use the results of the previous paragraph to predict the mean $\hat{u}$ and covariance $S$ of the components of the unit vector pointing from \gaia to the source in the reference frame at the epoch $T_0+\Delta T$. 

Finally, we accounted for the relativistic effect of aberration, which causes sources viewed by a moving observer to appear nearer to the apex of the observer's motion. The unit vectors $\vec{u}$ from the previous paragraph were derived using the location of the source on the sky in a frame that is stationary with respect to the Solar System barycentre, but aberration due to Gaia's orbit will causes the location of the source to vary between observations. \citet{klioner2003} states that the angular shift towards the apex is given by
\begin{align}
    \delta\theta =& \frac{1}{c}|\vec{v}_{\mathrm{G}}|\sin\theta\left[1+\frac{1}{c^2}(1+\gamma)w(\vec{x}_{\mathrm{G}})+\frac{1}{4}\frac{|\vec{v}_{\mathrm{G}}|^2}{c^2}\right] \nonumber \\
    &-\frac{1}{4}\frac{|\vec{v}_{\mathrm{G}}|^2}{c^2}\sin2\theta + \frac{1}{12}\frac{|\vec{v}_{\mathrm{G}}|^3}{c^3}\sin3\theta + O(c^{-4}), \label{eq:aberration}
\end{align}
where $\vec{v}_{\mathrm{G}}$ is the velocity of \gaia relative to the Solar System Barycentre at the epoch $T_0+\Delta T$ (obtained from the NASA JPL HORIZONS ephemeris calculator), $\theta$ is the angle of the source from the apex of \gaia's motion, $\gamma$ is a parameter in the Parameterized post-Newtonian formalism (we assume that General Relativity holds and thus $\gamma=1$), and $w(\vec{x}_{\mathrm{G}})$ gives the Solar System gravitational potential at \gaia's location. The angular shifts predicted by Eq. \ref{eq:aberration} are at most $20\;\mathrm{arcsec}$ for sources observed by \gaia. We note that due to \gaia's slow velocity ($30\;\mathrm{km}\;\mathrm{s}^{-1}$) and large distance from any massive bodies, almost all of the shift comes from the first term ($\delta\theta \approx \frac{1}{c}|\vec{v}_{\mathrm{G}}|\sin\theta$), with the remaining terms contributing at most $0.0005\;\mathrm{arcsec}$. We computed the angular shift due to aberration for every observation of every source and then rotated the unit vector $\vec{u}$ by that amount around the axis perpendicular to the velocity vector of \gaia and $\vec{u}$ using the rotation matrix $R$. When predicting observations of sources in Sec. \ref{sec:fractions} we used the first order expansion of Eq. 10 of \citet{klioner2003} that predicts that the unit vector $\vec{u}$ towards a source will be perturbed to $\vec{u}+\vec{v}_{\mathrm{G}}/c$, with the caveat that we re-normalise the resulting vector to ensure it retains unit normality. In subsequent paragraphs we will refer to the unit vector with aberration included as $\vec{u}$.

Suppose that at this epoch we have an uncertain estimate of the attitude of \gaia $\boldsymbol{\delta q}(\vec{a})\otimes\hat{\textbf{q}}$ parameterised by $\vec{a}$ with mean $\hat{\vec{a}}=0$ and covariance $P_{aa}$, then -- to first order in the uncertainties of the attitude and the unit vector in the reference frame -- the predicted unit vector in the body frame of \gaia has mean $A(\hat{\textbf{q}})\vec{u}$ and covariance $H_aP_{aa}H_a^T+A(\hat{\textbf{q}})SA^T(\hat{\textbf{q}})$, where $H_a = [\hat{\vec{u}}\times]$. The second term is simply the rotation of the uncertainty in the unit vector in the reference frame and we refer the reader to \citet{Markley2003} for details on the first term.

The key to the Multiplicative Extended Kalman Filter that we will describe in the following section is the measurement update step: the uncertain unit vector in the \gaia body frame pointing from \gaia to a source is predicted using our current estimate of \gaia's attitude, thus allowing us to use the unit vector deduced from the field-of-view, CCD and pixel of the detection as a measurement to constrain the attitude. 

\subsection{The Multiplicative Extended Kalman Filter}
\label{sec:mekf}

The multiplicative extended Kalman filter (MEKF) is a modification of the extended Kalman filter described in Sec. \ref{sec:kalmanfilters} that allows for part of the state -- the attitude of the spacecraft -- to be constrained to have unit normalisation. We stress that the overview of the MEKF presented here draws heavily from \citet{Markley2003}, who presented one of the seminal overviews of the MEKF, and \citet{Burton2017}, who adapted the MEKF for the case where the attitude needed to be determined based on noisy measurements of the vector pointing towards the Sun rather than measurements from a gyroscope or star-tracker.

The MEKF represents the true attitude quaternion of the spacecraft as
\begin{equation}
    \textbf{q}(t)=\boldsymbol{\delta q}(\vec{a}(t))\otimes \hat{\textbf{q}}(t),
\end{equation}
where $\hat{\textbf{q}}(t)$ is our expectation value of the attitude quaternion and $\boldsymbol{\delta q}(\vec{a}(t))$ encodes the uncertainty in that attitude (see Eq. \ref{eq:attitudeerror}). The state space of the MEKF is composed of three parts: the mean attitude quaternion $\hat{\textbf{q}}(t)$, the attitude error vector $\vec{a}(t)$ and the angular velocity vector $\vec{\omega}(t)$. The redundancy between the expectation value of $\vec{a}(t)$ and $\hat{\textbf{q}}(t)$ is removed by resetting $\vec{a}(t)$ to zero after each measurement update. The quantities that are explicitly tracked by the MEKF are thus the mean attitude and angular velocity vector $(\hat{\textbf{q}},\hat{\vec{\omega}})$ and the error covariance matrix
\begin{equation}
    P = \begin{bmatrix}P_{aa} & P_{a\omega} \\ P_{a\omega}^T & P_{\omega\omega} \end{bmatrix}.
\end{equation}
We assume that \gaia is subject to random zero-mean angular accelerations such that the dynamical equations are
\begin{align}
    \dot{\textbf{q}} &= \frac{1}{2}\begin{bmatrix}\vec{\omega} \\ 0 \end{bmatrix}\otimes \textbf{q} \\
    \dot{\vec{a}} &= \vec{\omega}-\hat{\vec{\omega}}+\vec{a}\times(\vec{\omega}+\hat{\vec{\omega}}) \\
    \dot{\vec{\omega}} &= \vec{w}, \label{eq:mekf_dynamics}
\end{align}
where $\vec{w}$ is drawn from a zero-mean multivariate normal distribution with the covariance matrix $Q=\sigma_{\mathrm{MEKF}}^2I_{3\times3}$. We note that in standard applications of the MEKF, Eq. \ref{eq:mekf_dynamics} would either have input from a gyroscope or include a model of the spacecraft dynamics in terms of torques and the moment of inertia. We have access to neither of these and thus rely on there being a sufficient number of variable star measurements that we capture the change in the angular velocity vector through time. Our estimates of these quantities evolve through time satisfying the equations
\begin{align}
    \dot{\hat{\textbf{q}}}(t) &= \frac{1}{2}\begin{bmatrix}\hat{\vec{\omega}} \\ 0 \end{bmatrix}\otimes\hat{\textbf{q}}(t) \\
    \dot{\hat{\vec{\omega}}}(t) &= 0 \\
    \dot{P}(t) &= FP(t) + P(t)F^T + GQG^T,
\end{align}
where
\begin{equation}
    F = \begin{bmatrix}\frac{\partial \dot{\vec{a}}}{\partial \vec{a}}\big\rvert_{\hat{\vec{a}},\hat{\vec{\omega}}} & \frac{\partial \dot{\vec{a}}}{\partial \vec{\omega}}\big\rvert_{\hat{\vec{a}},\hat{\vec{\omega}}} \\ \frac{\partial \dot{\vec{\omega}}}{\partial \vec{a}}\big\rvert_{\hat{\vec{a}},\hat{\vec{\omega}}} & \frac{\partial \dot{\vec{\omega}}}{\partial \vec{\omega}}\big\rvert_{\hat{\vec{a}},\hat{\vec{\omega}}} \end{bmatrix}
    = \begin{bmatrix}-[\hat{\vec{\omega}}\times] & I_{3\times3} \\ 0_{3\times3} & 0_{3\times3} \end{bmatrix},
\end{equation}
and
\begin{equation}
    G = \begin{bmatrix}\frac{\partial \dot{\vec{a}}}{\partial \vec{w}}\big\rvert_{\hat{\vec{a}},\hat{\vec{\omega}}} \\ \frac{\partial \dot{\vec{\omega}}}{\partial \vec{w}}\big\rvert_{\hat{\vec{a}},\hat{\vec{\omega}}}\end{bmatrix} = \begin{bmatrix} 0_{3\times3} \\ I_{3\times3} \end{bmatrix}.
\end{equation}
We refer the reader to \citet{Markley2003} and \citet{Burton2017} for the form of the derivations of these.

Analogously with Sec. \ref{sec:kalmanfilters}, we define $\hat{\textbf{q}}_{i|j}$, $\hat{\vec{\omega}}_{i|j}$ and $P_{i|j}$ to be our estimate of the mean attitude quaternion, mean angular velocity vector and covariance between the attitude error and angular velocity vectors at time $t_i$ given all of the measurements made at times up to and including time $t_j$. The MEKF iteratively calculates the quantities $\hat{\textbf{q}}_{i|i}$,  $\hat{\vec{\omega}}_{i|i}$ and $P_{i|i}$ at each time-step given all of the prior observations. Each time-step is split into a prediction stage and a measurement update stage.

At the prediction stage the dynamical equations above are solved with the expressions
\begin{align}
    \hat{\textbf{q}}_{i|i-1} &= \begin{bmatrix} \frac{\hat{\vec{\omega}}_{i-1|i-1}}{|\hat{\vec{\omega}}_{i-1|i-1}|}\sin\left(\frac{\delta t}{2}|\hat{\vec{\omega}}_{i-1|i-1}|\right) \\ \cos\left(\frac{\delta t}{2}|\hat{\vec{\omega}}_{i-1|i-1}|\right) \end{bmatrix}\otimes\hat{\mathbf{q}}_{i-1|i-1} \\
    \hat{\vec{\omega}}_{i|i-1} &= \hat{\vec{\omega}}_{i-1|i-1} \\
    P_{i|i-1} &= \Phi_{i|i-1}P_{i-1|i-1}\Phi_{i|i-1}^T + \delta t GQG^T,
\end{align}
where $\delta t = t_i - t_{i-1}$ and the state transition matrix is defined by $\Phi_{i|i-1} = \exp{(\delta t F_{i-1|i-1})}$. We exploited the structure of $F_{i-1|i-1}$ to avoid explicitly calculating the matrix exponential, as detailed in Appendix \ref{sec:matrixexponential}.

At the measurement stage we use the detected location of the star on the focal plane and the known location of the star on the sky to place a constraint on the attitude of \gaia. The unit vector of the star in the reference frame has mean $\hat{\vec{x}}_i$ and covariance $\Sigma_i$, while the unit vector of the detection in the body frame has mean $\hat{\vec{u}}_i$ and covariance $R_i$. The equations expressing this are
\begin{align}
    \hat{A} &= A(\hat{\textbf{q}}_{i|i-1}) \\
    \hat{\vec{m}}_i &= \hat{A}\hat{\vec{u}}_i \\
    H &= \begin{bmatrix} [\hat{\vec{m}}_i\times] & 0_3 \end{bmatrix} \\
    S &= H P_{i|i-1} H^T + \hat{A}\Sigma_i\hat{A}^T + R_i \\
    K &= P_{i|i-1}H^TS^{-1} \\
    \vec{c} &= \hat{\vec{u}}_i - \hat{\vec{m}}_i \\
    \begin{bmatrix} \hat{\vec{a}} \\ \Delta \hat{\vec{\omega}} \end{bmatrix} &= K\vec{c} \\
    \textbf{q}_{i|i} &= \boldsymbol{\delta q}(\hat{\vec{a}})\otimes \textbf{q}_{i|i-1} \label{eq:reset} \\
    \hat{\vec{\omega}}_{i|i} &= \Delta\hat{\vec{\omega}} + \hat{\vec{\omega}}_{i|i-1} \\
    P_{i|i} &= (I_{6\times6}-KH)P_{i|i-1} \\
    L_i &= -\frac{1}{2}\left(\vec{c}^TS^{-1}\vec{c} + \log|S|+2\log{2\pi}\right),
\end{align}
where the final line calculates the log-likelihood of the location of the detection given the location of the source on the sky and our prior estimate of the attitude. In practise, the rank deficiency of both $\Sigma_i$ and $R_i$ causes the inversion of $S$ to be numerically unstable. We solve this by discarding the rows and columns corresponding to the $x$ component of vectors and matrices in the body frame of \gaia, because all of the information gained during the measurement update is contained in the other two components. The redundancy between $\hat{\vec{a}}$ and $\hat{\textbf{q}}(t)$ is removed during each measurement update in Eq. \ref{eq:reset} by moving the mean rotation represented by a non-zero $\hat{\vec{a}}$ to the mean attitude quaternion $\hat{\textbf{q}}$. The total log-likelihood of the MEKF for a given set of values for the hyper-parameters is simply the sum of the individual log-likelihoods $L_i$ from each time-step.

The forward algorithm described above is sufficient to calculate the log-likelihood and to predict the attitude and angular velocity of \gaia at each time given all prior measurements, but it is possible to obtain a more precise estimate that additionally incorporates the measurements at later times through an altered version of the Rauch-Tung-Striebel smoothing algorithm \citep{Rauch1965} discussed in Sec. \ref{sec:kalmanfilters}. The required alterations were given in \citet{Kubelka2016}:
\begin{align}
    C_i &= P_{i|i}\Phi_{i+1|i}^TP_{i+1|i}^{-1} \\
    \begin{bmatrix} \hat{\vec{a}} \\ \Delta \hat{\vec{\omega}} \end{bmatrix} &=  C_i\begin{bmatrix} \boldsymbol{\delta q}^{-1}(\textbf{q}_{i+1|n}\otimes\textbf{q}_{i+1|i}^{-1})\\ \hat{\vec{\omega}}_{i+1|n}-\hat{\vec{\omega}}_{i+1|i}\end{bmatrix} \\
    \textbf{q}_{i|n} &= \boldsymbol{\delta q}(\hat{\vec{a}})\otimes \textbf{q}_{i|i} \\
    \hat{\vec{\omega}}_{i|n} &= \Delta\hat{\vec{\omega}} + \hat{\vec{\omega}}_{i|i} \\
    P_{i|n} &= P_{i|i} + C_i \left(P_{i+1|n} - P_{i+1|i}\right)C_i^T,
\end{align}
where the function $\boldsymbol{\delta q}^{-1}$ is the inverse of the Gibbs map in Eq. \ref{eq:attitudeerror}. Of the results in this subsection, only the generalisation of the measurement update to incorporate uncertainty in the known unit vector in the reference frame were original derivations of this work. The remaining results are a hybrid of \citet{Markley2003}, \citet{Burton2017} and \citet{Kubelka2016}.

\subsection{Application of the MEKF}
Our model of the attitude of \gaia has four free parameters -- the variance $\sigma_{\mathrm{MEKF}}^2$ of the random angular accelerations experienced by \gaia, the excess spread in the locations of the star on the focal plane $\sigma_p^2$ and the corrections to the CCD $\delta_c$ and pixel $\delta_c$ across-scan widths. If we propose a set of values for these parameters and provide a set of detections of \gaia sources then the MEKF algorithm described above will churn out a log-likelihood, and so we are able to carry out maximum-likelihood estimation of the values of those parameters. We opt to split the 22 months of \gaia DR2 into day-long chunks and perform maximum-likelihood optimisation separately in each day. We do this to make our estimation more robust to extreme adverse events; if \gaia's attitude rapidly changed in one short period then that could cause the value of $\sigma_{\mathrm{MEKF}}^2$ to be biased high. The MEKF is a recursive algorithm and so cannot be naively parallelised, and so splitting the optimisation into chunks also makes the optimisation more computationally tractable by allowing it to be spread across multiple CPUs. We performed the optimisation using the \textsc{scipy} implementation of the Nelder-Mead algorithm \citep{Gao2012}. The initial attitude and angular velocity vector was taken from the last time-point of the nominal scanning law prior to the first data-point in that day. We note that the nominal scanning law as provided by DPAC has a discontinuity on the 25 September 2014 ($\mathrm{OBMT} = 1326.7\;\mathrm{rev}$), which corresponds to the transition from the original nominal scanning law to one optimised for the `GAia Relativistic Experiment on Quadrupole' experiment \citep{deBruijne2010}. This transition created a discontinuity in the phase of both the precession and spin. Whilst in reality \gaia would smoothly transition to the updated scanning law, the nominal scanning file has an abrupt transition with the attitude changing by more than $100^{\circ}$ between two consecutive $10\;\mathrm{sec}$ time-steps. This transition occurs during the data-taking gap associated with the first decontamination of \gaia and so is not constrained by the variable star detections. We reset the MEKF at that time-point by resetting the attitude and angular velocity to that given by the nominal scanning law after the transition.

\begin{figure}
	\centering
	\includegraphics[width=1.0\linewidth,trim=0 0 0 0, clip]{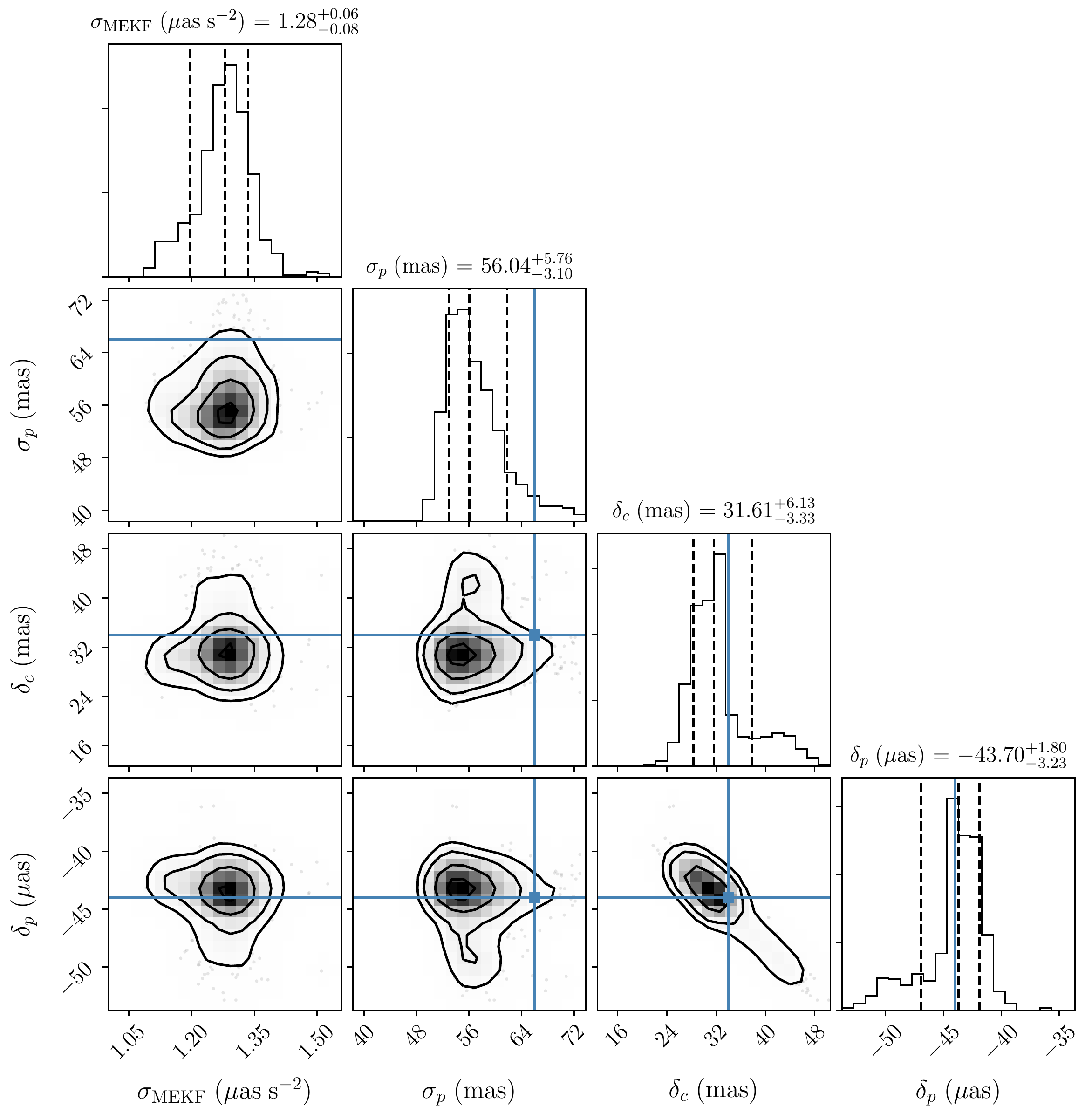}
	\caption{Corner plot of the hyperparameters of our \gaia attitude model, determining the random angular accelerations experienced by \gaia ($\sigma_{\mathrm{MEKF}}$), the excess error in the location of the stars ($\sigma_{p}$), and the along-scan corrections to the CCD $\delta_{c}$ and pixel $\delta_{p}$ widths. The blue lines indicate the values that the latter three of these hyperparameters took in our much simpler model of \gaia's attitude in \citetalias{PaperI}.}
	\label{fig:mekf_corner}
\end{figure}

In Fig. \ref{fig:mekf_corner} we show a corner plot of the maximum-likelihood values of $(\sigma_{\mathrm{MEKF}},\sigma_{p},\delta_{c},\delta_{p})$ from each of the 667 days. We estimate the values of our parameters from the (16,50,84)\% percentiles of this forest of maximum-likelihood values to be $\sigma_{\mathrm{MEKF}} = 1.28_{-0.08}^{+0.06}\;\mu\mathrm{as}\;\mathrm{s}^{-2}$, $\sigma_{p} =  56.04_{-3.10}^{+5.76}\;\mathrm{mas}$, $\delta_{c} =  31.61_{-3.33}^{+6.13}\;\mathrm{mas}$ and $\delta_{p} =  -43.70_{-3.23}^{+1.80}\;\mathrm{mas}$. The values of the latter three of these parameters (shown as blue lines in Fig. \ref{fig:mekf_corner}) are in excellent agreement with our findings in \citetalias{PaperI}, giving us faith in the results of these two independent methodologies, though the value of $\sigma_{p}$ is slightly smaller due to us mistakenly ignoring aberration in the previous work. We verified that if we neglected aberration in this work then we recovered the larger value of $\sigma_{p}$.

\begin{figure*}
	\centering
	\includegraphics[width=1.0\linewidth,trim=0 0 0 0, clip]{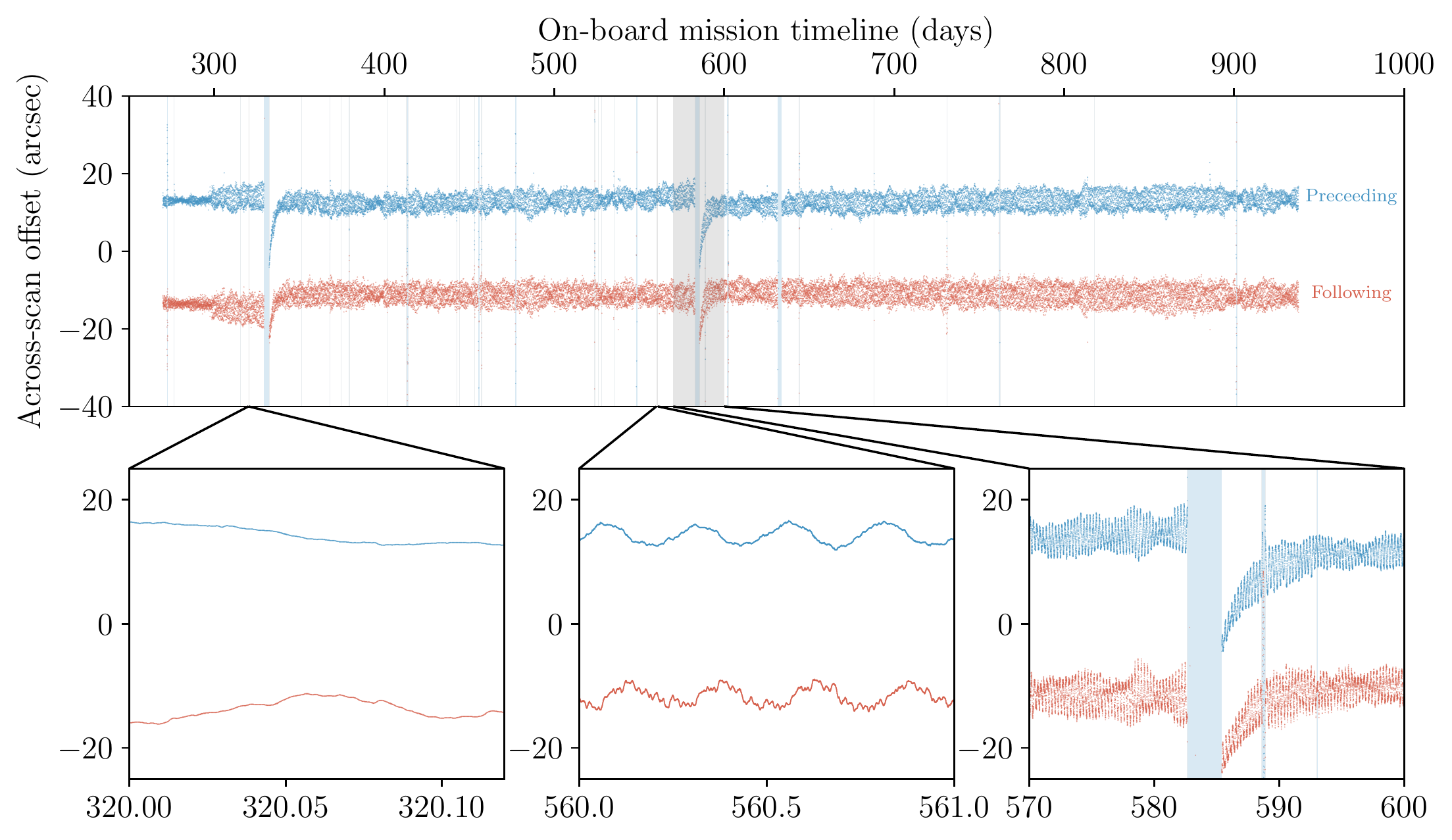}
	\caption{The across-scan distance between the locations of the two fields-of-view in our corrected scanning law and their locations in the nominal scanning law, in the frame of the nominal scanning law. This figure is comparable to Fig. 3 of \citetalias{PaperI}. The light blue bars in each panel indicate the gaps derived in Sec. \ref{sec:fractions}, during which there are no published detections of variable stars. \label{fig:attitude_ac}}
	

	\centering
	\includegraphics[width=1.0\linewidth,trim=0 0 0 0, clip]{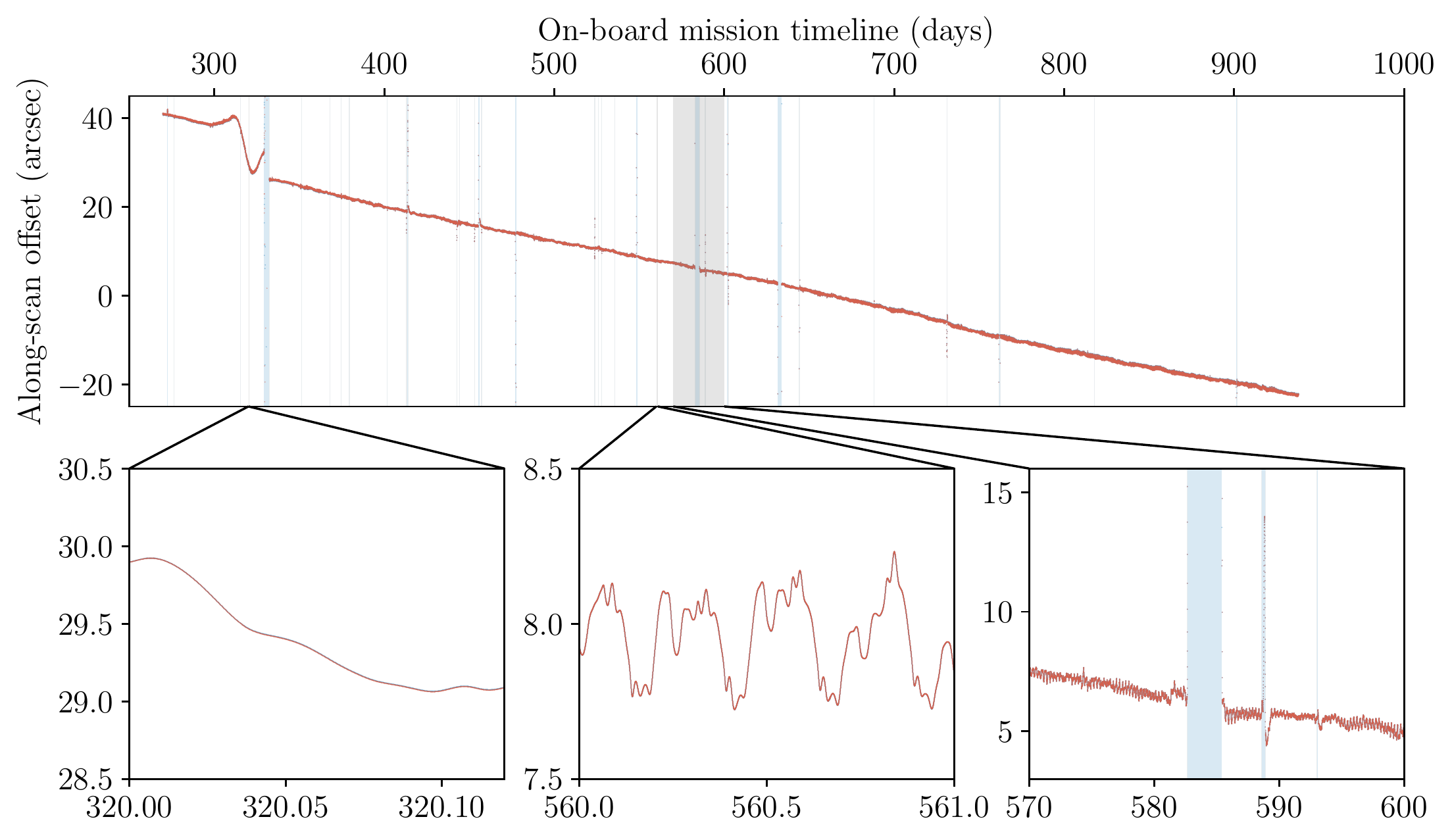}
	\caption{The along-scan distance between the locations of the two fields-of-view in our corrected scanning law and their locations in the nominal scanning law, in the frame of the nominal scanning law. The preceeding and following fields-of-view have identical along-scan offsets. \label{fig:attitude_al}}
	
\end{figure*}

Adopting the median values of the parameters as our fiducial values, we inserted the nominal scanning law time-points into the MEKF as time-points without a measurement update step and calculated the means and covariances of the attitude and angular velocity at those time-steps, employing the MEKF with the additional backwards smoothing step. In Figs. \ref{fig:attitude_ac} and \ref{fig:attitude_al} we show the across-scan and along-scan distance between the locations of the two fields-of-view in our corrected scanning law and their locations in the nominal scanning law, in the frame of the nominal scanning law. The across-scan offsets of the two fields-of-view have medians ($12.9$ and $-11.4\;\mathrm{arcsec}$ respectively) that are significantly different from zero, indicating either a long-term difference between the two scanning laws or that our value of $\Delta_f$ is too small by approximately $12\;\mathrm{arcsec}$. \textbf{Fig. \ref{fig:attitude_ac} shows significant differences to the equivalent figure of \citetalias{PaperI}, because in that previous work we failed to include relativistic aberration when deriving the scanning law, causing our model in that paper to fit the aberration and thus resulting in periodic structure on the $63\;\mathrm{day}$ precession period.} There is, however, periodic behavior in the across-scan offsets on the 6 hour spin period of \gaia. These are projections of small differences in \gaia's angular velocity between the two scanning laws that cause the angular offset to grow and shrink as \gaia rotates. \footnote{An analogy can make this clearer. Imagine we release a satellite on a circular orbit about the Earth and at the same time release a second satellite on a slightly inclined but otherwise identical orbit. The distance between these two satellites will oscillate with a period that is identical to the period of their orbit. Similarly, small offsets in the estimated angular velocity of \gaia between the nominal scanning law and our derived scanning law cause the angular offset between the two to oscillates with \gaia's rotation periods.} The along-scan offsets of the two fields of view are identical, which is a geometric constraint imposed by our model. Fig. \ref{fig:attitude_al} shows an offset that changes linearly with time, except from the period shortly after $\mathrm{OBMT}=300\;\mathrm{days}$ where \gaia transitions from the Ecliptic Pole Scanning Law to the Nominal Scanning Law (the nominal scanning law published by DPAC has a discontinuity at this transition). During periods where none of the variable stars have published detections our scanning law deviates significantly from the nominal scanning law, because we have not directly modelled the precession of \gaia's spin axis. However, this is not a significant issue, because detections taken during these periods will not have contributed to the \gaia DR2 data products. We will make our scanning law publicly available upon acceptance of this manuscript.

\section{Efficiency of obtaining useful detections from observations}
\label{sec:fractions}

If we had perfect knowledge of \gaia's scanning law then we could predict the number of times that any star would have transited across either of the two fields-of-view, which we will term to be a \textit{predicted observation}. Not every observation of a star results in a measurement in the \gaia DR2 variable star epoch table, which we will term to be a \textit{published detection}. There are several reasons why a predicted observation might not result in a published detection:
\begin{enumerate}
    \item \gaia experienced events which resulted in some detections not being of sufficient quality for publication in DR2 (e.g. the decontamination procedures or micro-meteoroid impacts, see \citealp{Gaia2016}). We term these periods to be \textit{gaps} because no star of any magnitude has any published detections from during these periods. Part of the data taken during DR2 gaps may appear in subsequent data releases.
    \item An observation only results in a published detection with some probability which depends on the properties of the source. This may be because the occurrence of a detection is probabilistic (for instance, at the faint magnitude limit photon shot noise will result in stars being detected on only a fraction of their observations) or because of quality cuts made by the \gaia DPAC.
    \item \gaia has a limited capacity to assign windows to stars and store and down-link the resulting scientific data, and so if densely-populated regions are being scanned then only some fraction of the predicted observations will result in data that makes it to the ground. The retention behaviour is decided by magnitude bin (see Tab. \ref{tab:spbins}), with some bins being prioritised to ensure good calibration across the entire magnitude range.
\end{enumerate}

The objective of this section is to rigorously identify the gaps in \gaia data-taking by exploiting the DR2 variable star epoch photometry. We attempted this in \citetalias{PaperI} by looking for periods longer than 1\% of a day where none of the variable stars had a published detection, but there were several possible issues with this approach. First, this approach cannot distinguish between a period with no published detections and a period where no variable stars were observed because \gaia was scanning a sparse region of the sky. Second, our choice of 1\% of a day was an attempt to mitigate the first problem and it is likely that there are gaps shorter than this. Third, gaps are not the only reason that an observation might not result in a detection, as mentioned above.

Our novel methodology in this work is based on the idea that a run of non-detections could be due to a gap or it could be due to those stars having a low probability that an observation would result in a published detection. By pairing the published detections with the predicted observations that failed to result in published detections and weighing the possibility of a gap against the possibility of low detection efficiencies, we are able to search for gaps in a way that is robust to the density of variable stars on the sky and to the low detection efficiency of faint stars. We model the probability of a source with magnitude $G$ having a published detection at the predicted observation time $t$ as $A(t)B(G,t)$, where $A(t)$ takes binary values and represents gaps that affect all magnitudes equally while $B(G,t)$ can take any value in $(0,1)$ and represents any other effects that can cause an observation to not yield a published detection. We assume that the magnitude-dependency of $B(G,t)$ is piecewise-constant such that at time $t$ it takes a different value in each of the star packet magnitude intervals (see Tab. \ref{tab:spbins}), because one of the most important drivers of a time-varying detection probability is crowding and the impact of crowding changes between each star packet magnitude interval.

In Sec. \ref{sec:predictingobs} we discuss the preparation of the data and the prediction of the times of observation.  We describe our discrete hidden Markov model for $A(t)$ in Sec. \ref{sec:discrete} and our Extended Kalman Filter model for $B(G,t)$ in Sec. \ref{sec:continuous}. The full Bayesian specification of this problem would require us to have a posterior over millions of parameters (one discrete and one continuous hidden state at each epoch measurement in addition to the tens of hyperparameters) and so we opt for an iterative maximum likelihood approach which we describe in Sec. \ref{sec:maximumlikelihood}. 

\subsection{Predicting observations and data cleaning}
\label{sec:predictingobs}

We use the 550,737 variable stars in the \gaia DR2 epoch photometry tables. We extract the times of observation at \gaia from the \textsc{transit\_id} and approximate the error in the epoch $G$ magnitude from the reported epoch flux and flux error. For each source in the epoch photometry table we then predict when it would have been observed during the 22 months of \gaia DR2, following the methodology laid out in \citetalias{PaperI} and \citetalias{PaperII} and using the scanning law derived in Sec. \ref{sec:scanninglaw}. The resulting 19,685,650 predicted observations were then matched to the published detections with a window of $1\%$ of a day, with the predicted observations being assigned a value of one if they are matched with a published detection and zero otherwise.

Our methodology is magnitude dependent and so we must assign a magnitude to each predicted observation. In the case of those with matching published detections we take the magnitude estimate and error reported at that epoch, but for the non-detections we must infer the magnitude at those epochs. The na\"ive approach would be to assign the mean $G$ band magnitude from the \gaia source table to be the magnitude at each non-detection, but that ignores that these stars have been identified to be variable stars and thus the magnitude will significantly change between epochs. The best way to proceed would be to have theoretical models that predict the light-curves of different types of variable star objects, fit those model light-curves to the published detections in a Bayesian way, and then marginalise over those fits to obtain a predicted measurement at each of the predicted observations without a published detection. However, this is far beyond the scope of this work and would likely only make a marginal difference to our results. We will be combining the measurements from many stars into magnitude bins and thus a few observations crossing the border into an adjacent magnitude bin shouldn't overly bias our results. We therefore decided to infer these values using Gaussian process regression, to which we gave an introduction in Sec. \ref{sec:gaussianprocesses}. The reader should view our use of Gaussian process regression in this instance as a mathematically convenient curve-fitting scheme that can account for uncertainty in the individual magnitude measurements.

\begin{figure}
	\centering
	\includegraphics[width=1.0\linewidth,trim=0 0 0 0, clip]{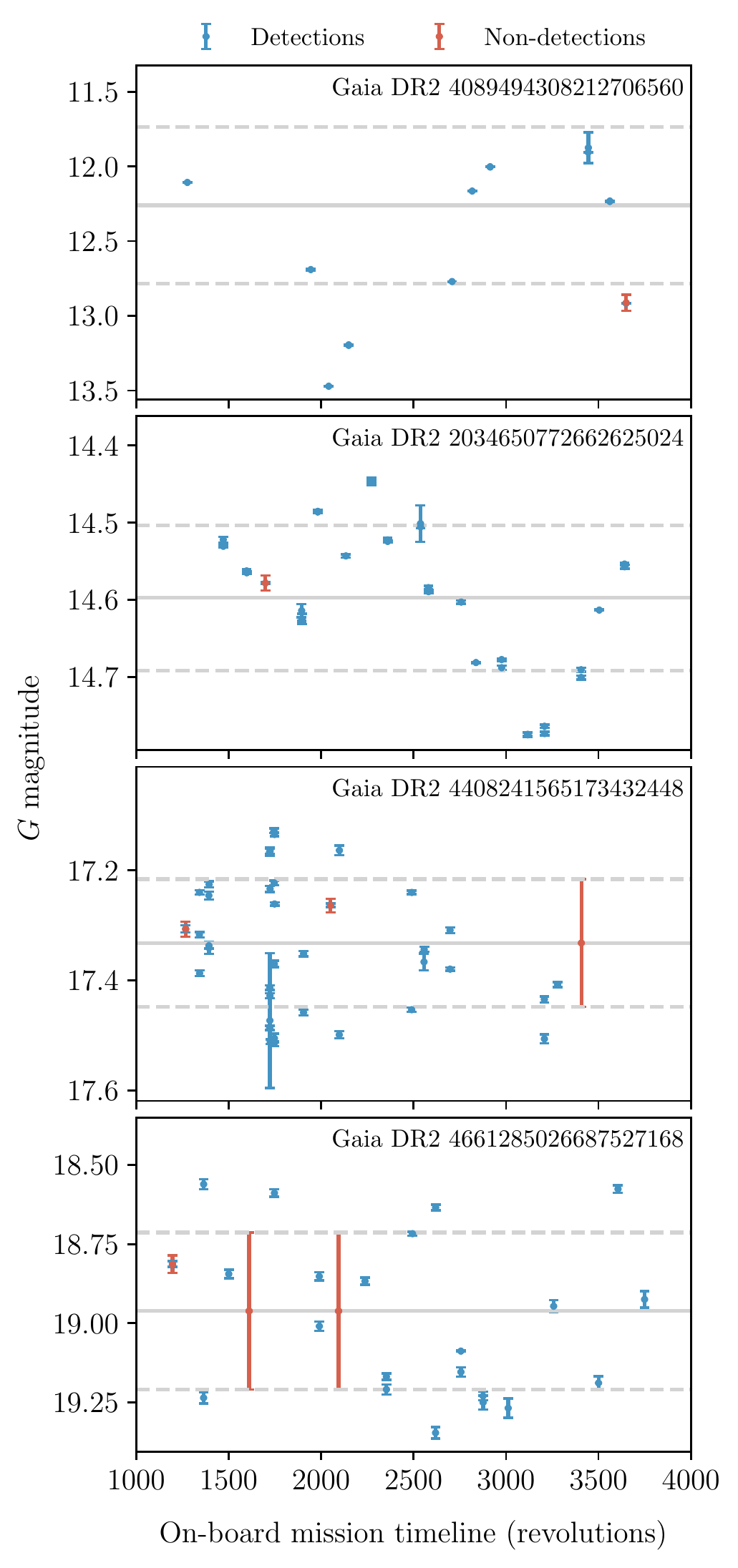}
	\caption{Examples of the results from our use of a Gaussian Process to impute the magnitudes of the variable stars at the epochs at which they were predicted to be observed but that did not result in a published detection in the \gaia DR2 epoch photometry tables. The solid and dashed lines indicate the weighted mean and one sigma regions of the published detections of each star.}
	\label{fig:gaussianprocess}
\end{figure}

We assume that the time-evolution of the magnitude of each star can modelled as the realization of a Gaussian process.
We have measurements of the magnitude $G(t_{i})$ and magnitude variance $\sigma^{2}(t_{i})$ at times $\{t_{i}\}_{i = 1}^{n}$ and wish to predict the magnitude at any time $t_{*}$.
We assume that the mean is constant, and that the covariance kernel is given by Eq.~\ref{eq:gp_cov_kernel}.
We fix the mean and signal variance according to Eq.~\ref{eq:gp_sample_mean} and~\ref{eq:gp_sample_variance}.
Though the actual period distribution of the published variables stars ranges from hours till hundreds of days, we fix the length scale to $l = 1$~day as a very rough `typical' value.
This has the consequence that a non-detection occurring soon before or soon after a detection will be assigned a magnitude close to that of the detection, whereas a non-detection occurring long before or long after a detection will be assigned a magnitude close to the sample mean.
The scanning law causes most stars observed in the preceding field-of-view to be observed two hours later by the following field-of-view and thus observations come in pairs. Our inference scheme is therefore most effective in the fairly common case that one of these observations did not result in a detection. We give examples of the use of Gaussian process regression to infer the magnitudes of non-detections in Fig.~\ref{fig:gaussianprocess}.

We applied this formalism to all 550,737 of the stars with epoch photometry. We then used the magnitude at each predicted observation to group them across all stars by magnitude into the bins given in Table \ref{tab:spbins}, and further sorted these observations by time. If there is a matching published detection then the magnitude used for the grouping is the one measured by \gaia, otherwise it is the magnitude infilled by our Gaussian Process as described above. 

In summary, we have magnitude bins $i=1,\dots,19$ each with a time-series $t_j^i$ of predicted observations $j=1,\dots,M^i$ each with a flag $k_j^i$ which indicates whether there is a matching published detection ($k_j^i=1$) or not ($k_j^i=0$). 

\begin{table}
\centering
\caption{The magnitude bins used on-board by \gaia to determine priority for deletion if the downlink bandwidth is exhausted and which we used to bin the epoch measurements in this work. 
The first three columns of this table are a reproduction of  \citealp[Table 1.10 Sec. 1.3.3 of][]{2018gdr2.reptE...1D}. }
\begin{tabular}{lcccc}
\hline Packet & Magnitude range & Deletion (\%) & $l\;(\mathrm{day})$ & $\varepsilon\;(\mathrm{day}^{-1})$ \\
\hline SP1-1 & (5.00,13.00) & 0 & 150.49 & 1.53  \\
SP1-2 & (13.00,16.00) & 0 & 1.57 & 1.58  \\
SP1-3 & (16.00,16.30) & 1 & 9.65 & 1.48  \\
SP1-4 & (16.30,17.00) & 1 & 1.78 & 1.52  \\
SP1-5 & (17.00,17.20) & 2 & 7.01 & 1.52  \\
SP1-6 & (17.20,18.00) & 2 & 2.15 & 1.52  \\
SP1-7 & (18.00,18.10) & 2 & 11.11 & 1.51  \\
SP1-8 & (18.10,19.00) & 2 & 2.01 & 1.44  \\
SP1-9 & (19.00,19.05) & 2 & 7.02 & 1.30  \\
SP1-10 & (19.05,19.95) & 7 & 0.28 & 1.21  \\
SP1-11 & (19.95,20.00) & 2 & 7.87 & 1.00  \\
SP1-12 & (20.00,20.30) & 13 & 1.39 & 0.99  \\
SP1-13 & (20.30,20.40) & 12 & 2.32 & 1.08  \\
SP1-14 & (20.40,20.50) & 13 & 2.21 & 1.07  \\
SP1-15 & (20.50,20.60) & 28 & 2.51 & 1.02  \\
SP1-16 & (20.60,20.70) & 28 & 3.01 & 0.97  \\
SP1-17 & (20.70,20.80) & 28 & 8.64 & 0.84  \\
SP1-18 & (20.80,20.90) & 28 & 19.67 & 0.73  \\
SP1-19 & (20.90,21.00) & 24 & 15.10 & 0.63  \\     \hline
\end{tabular}
\label{tab:spbins}
\end{table}

\begin{figure*}
	\centering
	\includegraphics[width=1.0\linewidth,trim=0 0 0 0, clip]{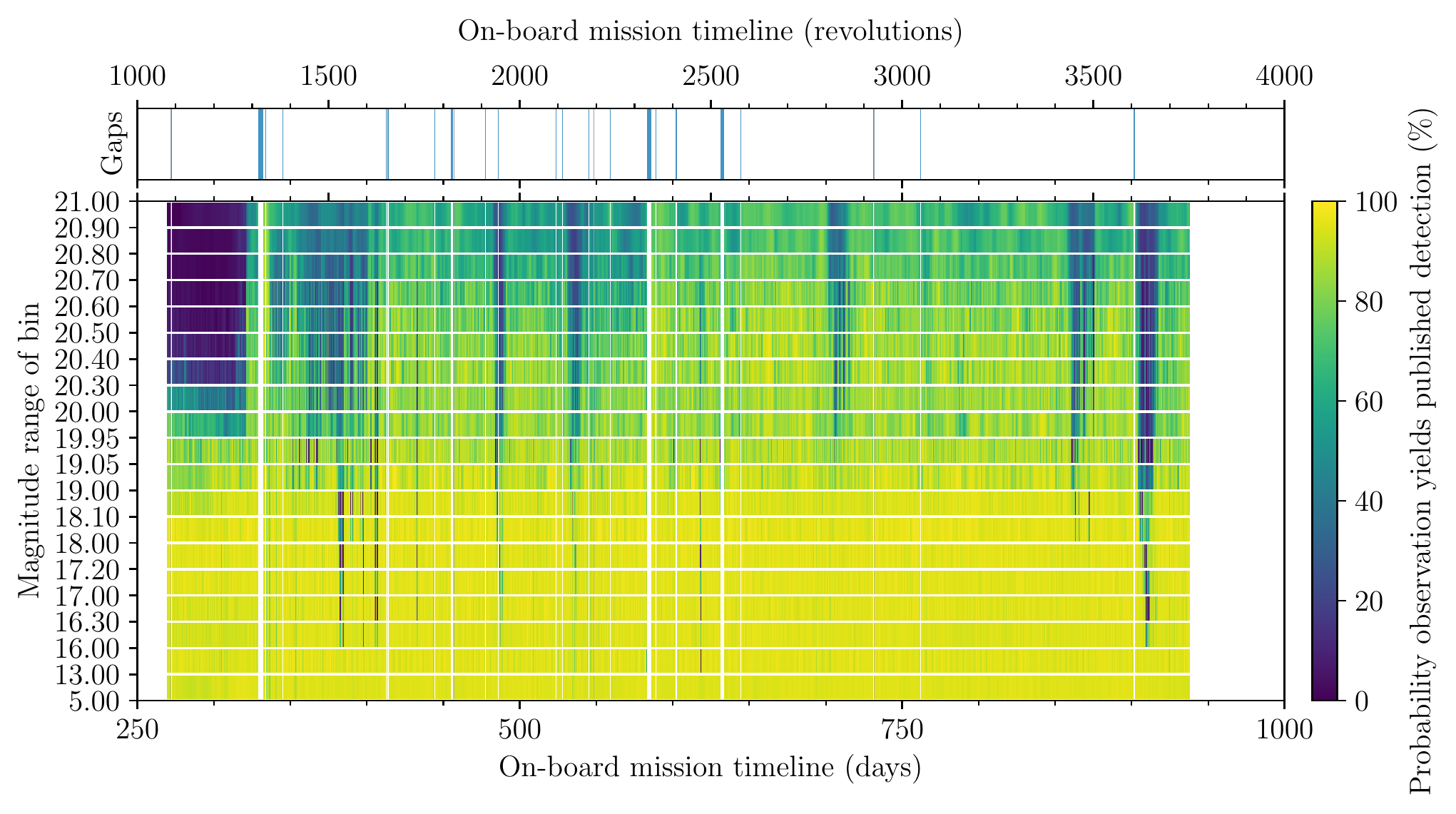}
	\caption{Illustration of our final gaps and probabilities of detection. \textbf{Top:} The inferred gaps where no \gaia observations resulted in published detections are shown in blue. \textbf{Bottom:} For each of the magnitude bins used to define \gaia star packets, the colour map shows the probability that a \gaia observation made at that time would result in a published detection.}
	\label{fig:fractions}
\end{figure*}

\subsection{Hidden Markov Model for gaps}
\label{sec:discrete}

We will model the occurrence of gaps with a Hidden Markov Model, which we introduced in Sec. \ref{sec:hiddenmarkov}. In this application the hidden state $X_i$ at time $t_i$ indicates whether a gap is occurring ($X_i=0$) or not ($X_i=1$), and the observation $Y_i$ reports whether or not the predicted observation of a variable star at time $t_i$ has yielded a published detection ($Y_i=1$) or not ($Y_i=0$). We will assume that no detections are possible during gaps 
\begin{equation}
    P(Y_i=1|X_i=0) \equiv 0 \Rightarrow P(Y_i=0|X_i=0) \equiv 1,
\end{equation}
and that at each step there is some known probability $p_i$ that a published detection occurs if we are outside of a gap
\begin{equation}
    P(Y_i=1|X_i=1) \equiv p_i \Rightarrow P(Y_i=0|X_i=1) \equiv 1-p_i.
\end{equation}
The value of $p_i$ will depend on the magnitude of the star being observed at time $t_i$ and will be an output of the model in the following section.

The transition between states will be governed by the probability $q_i$ that a gap will begin at step $i$ and the probability $r_i$ that a gap will end at step $i$,
\begin{align}
    P(X_i=0|X_{i-1}=1) &\equiv q_i &\Rightarrow P(X_i=1|X_{i-1}=1) &\equiv 1-q_i, \\
    P(X_i=1|X_{i-1}=0) &\equiv r_i &\Rightarrow P(X_i=0|X_{i-1}=0) &\equiv 1-r_i.
\end{align}
The predicted observation times are not uniformly spaced and so we assume that
\begin{align}
    q_i &= \frac{1}{2}-\frac{1}{2}\exp{-\left(\frac{t_i-t_{i-1}}{q}\right)}, \nonumber \\
    r_i &= \frac{1}{2}-\frac{1}{2}\exp{-\left(\frac{t_i-t_{i-1}}{r}\right)},
\end{align}
where $q$ and $r$ are the length-scales over which non-gaps and gaps persist. These expressions have the properties that for $t_i-t_{i-1}\ll 1$ the probability of the state changing is zero and that if $t_i$ is a long time after $t_{i-1}$ then the two states are entirely uncorrelated ($X_i$ is equally likely to zero or one regardless of the state $X_{i-1}$).

\subsection{Continuous state space modelling}
\label{sec:continuous}

We will model the probability of an observation at time $t_i$ resulting in a published detection \textit{if a gap is not occurring} with an Extended Kalman Filter in each of the magnitude bins. We have observations $y_i$ at predicted observation times $t_i$ at which a star has a published detection ($y_i=1$) or not ($y_i=0$). We assume that the probability of a published detection is given by
\begin{equation}
    \operatorname{P}(y_i|x_i)=\begin{cases}
\Phi(x_i) &\text{if $y_i = 1$,}\\
1-\Phi(x_i) &\text{if $y_i = 0$,}
\end{cases}
\label{eq:observations}
\end{equation}
where $X_i\in \mathbb{R}$ is the hidden state and $\Phi(X_i)$ maps $X_i$ to $(0,1)$. Throughout this section we will use the notation $\phi(x|\mu,\sigma^2)$ and $\Phi(x|\mu,\sigma^2)$ to denote the probability density and cumulative density functions of a normally-distributed random variable with mean $\mu$ and variance $\sigma^2$, where if the givens are omitted -- such as in Eq. \ref{eq:observations} -- we are referring to the standard Normal distribution ($\mu=0,\sigma^2=1$). For convenience, we give the explicit forms of these functions,
\begin{align}
    \phi(x|\mu,\sigma^2) &= \frac{1}{\sqrt{2\pi\sigma^2}}e^{-\frac{x-\mu}{2\sigma^2}},\\
    \Phi(x|\mu,\sigma^2) &= \frac{1}{2}\left(1+\operatorname{erf}\left(\frac{x-\mu}{\sqrt{2\sigma^2}}\right)\right).
\end{align}
The use of $\Phi(x)$ in Eq. \ref{eq:observations} is solely to allow us to work with state space variables in $(-\infty,+\infty)$ and then map them to $(0,1)$, and the choice of this function over other functions that can carry out this mapping (such as a sigmoid function) was made to simplify integrals later in this section.

We assume that the state variables $X_i$ are related to each other through a variation on an Extended Kalman filter 
\begin{align}
    x_{i} &= F_ix_{i-1} + w_i, \\
    F_i &= e^{-\tau_i/l},  \\
    w_i &\sim\operatorname{Normal}\left(0,(1-e^{-2\tau_i/l})\varepsilon^2\right), \\
    \tau_i &= t_i-t_{i-1},
\end{align}
where $l$ and $\varepsilon^2$ are the length-scale and variance of the process. These equations can be equivalently expressed through the probability density function of the next state conditioned on the previous state,
\begin{equation}
    \operatorname{P}(x_{i}|x_{i-1}) = \phi\left(x_{i}\bigg\lvert e^{-\tau_i/l}x_{i-1},\left(1-e^{-2\tau_i/l}\right)\varepsilon^2\right).
\end{equation}
We note that this process is mean-reverting: as the time between observations grows large the conditional expectation of the next observation tends to zero.

\begin{figure}
	\centering
	\includegraphics[width=1.0\linewidth,trim=0 0 0 0, clip]{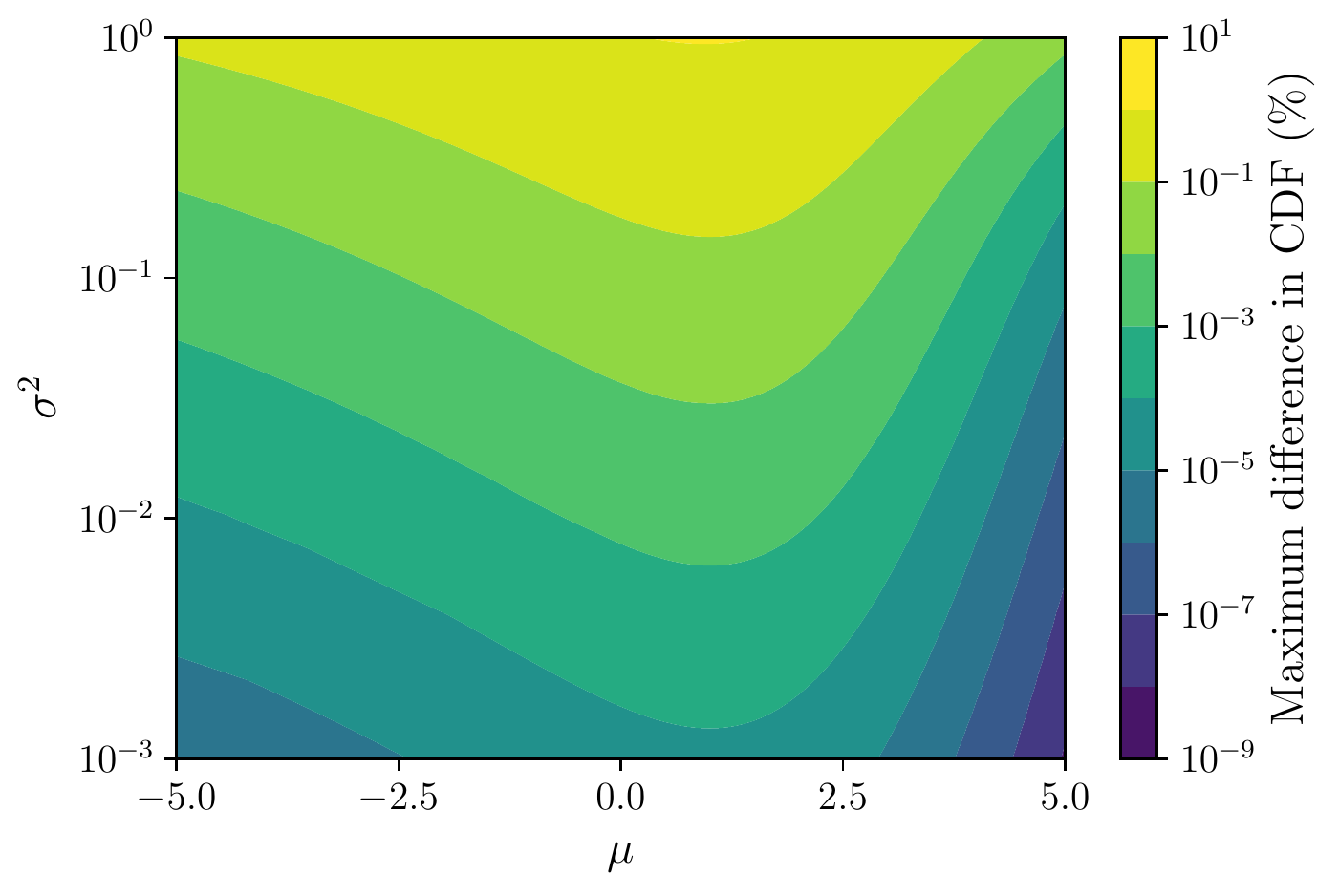}
	\caption{The maximum percentage difference between the cumulative densities of the function $\Phi(x)\phi(x|\mu,\sigma^2)$ and the Gaussian approximation described in the text.}
	\label{fig:approximation}
\end{figure}

Throughout this section we will encounter expressions of the form $f(x)=\Phi(x)\phi(x|\mu,\sigma^2)$ and will need to approximate $f(x)\approx A\phi(x|m,s^2)$. Denote the $j$-th moment of $f(x)$ about zero by $J_j=\int_{-\infty}^{+\infty}x^jf(x)\mathrm{d}x$, then we set $A=J_0$ to ensure the integral of $f(x)$ is preserved and further choose $m\approx J_1/J_0$ and $s^2\approx J_2/J_0-m^2$, i.e. we set the mean and variance of the approximation to be the mean and variance of the normalised $f(x)$. The moments are given by the expressions
\begin{align}
    J_0 &= \frac{1}{2}\left(1+\operatorname{erf}\left(\frac{\mu}{\sqrt{2}\sqrt{1+\sigma^2}}\right)\right), \\
    J_1 &= \mu J_0 + \sigma^2\phi(0|\mu,1+\sigma^2), \\
    J_2 &= \sigma^2 J_0 + \mu J_1 + \frac{\mu\sigma^2}{1+\sigma^2}\phi(0|\mu,1+\sigma^2).
\end{align}
For $(\mu,\sigma^2)=(0,1)$ the difference between the normalised cumulative density functions is less than 1\% everywhere, while for $(\mu,\sigma^2)=(0,0.1)$ this difference is less than 0.05\%, as shown in Fig. \ref{fig:approximation}. For our purposes $\sigma^2$ is typically much less than unity and so this approximation is excellent.

We need to obtain both the likelihood of the hyper-parameters $l$ and $\varepsilon^2$ and the best estimate of the states ${X_i}_{i=1}^{n}$ given the observations ${Y_i}_{i=1}^{n}$. The reason our model cannot be expressed as a conventional Extended Kalman Filter is that our observations are not continuous and thus we cannot simply differentiate the observation model with respect to the state space parameters. We instead follow the formalism laid out in Ch. 4 of \citet{Fraser2008}.

Denote by $\mu_{i|j}$ and $\sigma_{i|j}^2$ the mean and variance of the estimate of the state at time $t_i$ given all of the observations that occurred up to and until time $t_j$. We initialise the algorithm by specifying a weak prior on the first state $P(x_1)=\phi(x_1|-3,6)$. In the Forward part of the algorithm we iteratively forecast the next state given all previous observations
\begin{align}
    P&(x_{i+1}|y_1,\dots,y_i) = \int P(x_{i+1}|x_i)P(x_i|y_1,\dots,y_i) \mathrm{d}x_i \\
    &= \int \phi(x_{i+1}|e^{-\tau_i/l}x_i,\left(1-e^{-2\tau_i/l}\right)\varepsilon^2)\phi(x_i|\mu_{i|i},\sigma^2_{i|i}) \mathrm{d}x_i \\
    &= \phi(x_{i+1}|\mu_{i+1|i},\sigma^2_{i+1|i}),
\end{align}
using the fact that the convolution of two normal distributions is another normal distribution, where
\begin{align}
    \mu_{i+1|i} &= \mu_{i|i}e^{-\tau_i/l} \\
    \sigma_{i+1|i}^2 &= \sigma_{i|i}^2e^{-2\tau_i/l}+\varepsilon^2\left(1-e^{-2\tau_i/l}\right).
\end{align}
We can then obtain the conditional state distribution and probability of the observation as
\begin{align}
    P&(x_{i+1}|y_1,\dots,y_{i+1})P(y_{i+1}|y_1,\dots,y_i) \nonumber \\
    &= P(y_{i+1}|x_{i+1})P(x_{i+1}|y_1,\dots,y_i) \nonumber \\
    &= \Phi(x_{i+1})\phi(x_{i+1}|\mu_{i+1|i},\sigma^2_{i+1|i}) \nonumber \\
    &\approx L_{i+1|i}\phi(x_{i+1}|\mu_{i+1|i+1},\sigma^2_{i+1|i+1}) \label{eq:moment_matching}
\end{align}
where we obtain $L_{i+1|i}$, $\mu_{i+1|i+1}$ and $\sigma^2_{i+1|i+1}$ through the moment-matching approximation discussed above. Note that Eq. \ref{eq:moment_matching} is only true for the case that $y_{i+1}=1$. If $y_{i+1}=0$ then $P(y_{i+1}|x_{i+1})=1-\Phi(x_{i+1})$ and the expressions above can be trivially adjusted. The total likelihood $L$ is then given by
\begin{equation}
    L = L_1\prod_{i=1}^{n-1}L_{i+1|i}
\end{equation}
where $L_1$ can be computed by replacing $P(x_{i+1}|x_{i})$ by the prior $P(x_1)$ in the equations above. Having approximated all of the conditional probability distributions in the forward algorithm as normal distributions, the backwards smoothing filter described in Sec. \ref{sec:kalmanfilters} can be applied to compute $\mu_{i|n}$ and $\sigma_{i|n}^2$, noting $F_i=\exp(-\tau_i/l)$.

If there is known to be a gap at time $t_i$, then we adjust the expressions above by stating $P(y_i|x_i)=1$.

\subsection{Estimating the hyper-parameter likelihood and discussion}
\label{sec:maximumlikelihood}

Our model for the gaps has two free parameters $q$ and $r$ and our model for the detection efficiency has two free parameters $l_i$ and $\varepsilon_i^2$ in each magnitude bin $i$, for a total of thirty eight free parameters. The run-time of both algorithms is linear in the number of predicted observations and thus it would be prohibitively expensive to simultaneously optimise all parameters. We opt instead to alternate between optimising the free parameters in the gaps model and optimising the free parameters in the detection efficiency model, using the maximum likelihood gaps or efficiencies from the last optimisation of each in the next optimisation of the other.

Our final time-scales for the persistence of non-gaps and gaps were $q=1.216$ and $r=0.043$ days. If the time between the current state and the next is equal to the persistence time-scale then the probability of the next state being the same as the current state is 68.4\%, while if the time difference is twice the persistence time-scale then the probability of remaining in the same state is only $56.8\%$. Our final values for $q$ and $r$ tell us that gaps are not frequent and tend to be short-lived. The final values for the time-scale $l$ and variance $\varepsilon^2$ of the detection efficiency process in each magnitude bin are given in Tab. \ref{tab:spbins}. These describe the time-scales over which the detection efficiencies appear to vary and the scale of those variations. We illustrate the final gaps and detection efficiencies in Fig. \ref{fig:fractions}.

Fig. \ref{fig:fractions} shows rich structure which can be attributed to known effects. The drops in detection probability around $\mathrm{OBMT}$ $1925\;\mathrm{rev}$, $2150\;\mathrm{rev}$, $2825\;\mathrm{rev}$, $3475\;\mathrm{rev}$ and $3575\;\mathrm{rev}$ are due to periods when \gaia was frequently scanning across the Galactic plane. The initial period of low detection probability for sources fainter than $G=20.3$ is due to the faint-end threshold for the SkyMapper CCDs to register a detection \citep[][]{deBruijne2015} originally being set to $G=20.3$ . This threshold was changed to $G=21.0$ on 15 September 2014 and decreased to its final value of $G=20.7$ on 27 October 2014 \citep[see Sec. 1.3.3 of][]{2018gdr2.reptE...1D}. These threshold changes do not appear as crisp edges in this plot because the SkyMapper-estimated $G$ magnitude is not as accurate as the calibrated measurement made by the astrometric CCDs. We identified a total of 206 gaps, an increase from the 94 gaps identified in \citetalias{PaperI}. The gaps which are clearly visible in Fig. \ref{fig:fractions} were primarily caused by the mirror decontaminations and subsequent refocusings, station-keeping maneuvers and micro-meteoroid impacts.

\section{Using our scanning law to predict Gaia observations and the probability of them resulting in detections}
The results of our series of papers rely heavily on predicting when \gaia observed a location on the sky, given the scanning law. Being able to predict \gaia observations would be useful in other science cases beyond inferring \gaia's completeness, and so we have produced a \textsc{Python} module \textsc{scanninglaw} (\url{https://github.com/gaiaverse/scanninglaw}) based on the \textsc{dustmaps} package by \citet{Green2018} and subsequent \textsc{selectionfunctions} package from \citetalias{PaperII}. This enables the user to ask the question `At what times could \gaia have observed my star in the time frame of DR2 and what is the probability that each observation was successfully processed?'. This is demonstrated by determining when the fastest main-sequence star in the Galaxy \citep[S5-HVS1,][]{Koposov2020} would have been observed in \gaia DR2. This Python package has options to download and use the DPAC nominal scanning law or the one derived in Sec. \ref{sec:scanninglaw} of this work.

\begin{lstlisting}[language=Python]
import scanninglaw.times as times
from scanninglaw.source import Source

dr2_sl = times.dr2_sl(version='cog')
s5_hvs1 = Source('22h54m51.68s',
                 '-51d11m44.19s',
           photometry={'gaia_g':16.02},
           frame='icrs')
obs_t, obs_f = dr2_sl(s5_hvs1, 
                return_fractions=True, fov=1)
print('obs times FoV1:', obs_t)
print('obs fraction FoV1:', obs_f)

>> times FoV1: [[1714.17326, ..., 2325.42742]]
>> fracs FoV1: [[0.94585, ..., 0.95110]]
\end{lstlisting}

\section{Conclusions}
\label{sec:conclusion}

The completeness of the \gaia catalogues is heavily dependent on the status of \gaia through time. If there is a gap in scientific operations, a drop in the detection efficiency or \gaia deviates from the commanded scanning law, then stars will miss out on potential detections and thus be less likely to make it into the \gaia catalogues. The \gaia mission will take hundreds of epoch astrometric, photometric and spectroscopic measurements of billions of stars, which will implicitly encode the status of \gaia throughout the mission. In this work we have laid the groundwork for the future exploitation of these massive time-series by developing novel methodologies to infer the orientation of \gaia and the gaps and detection efficiencies from time-series of \gaia detections. We have applied these methodologies to the \gaia DR2 variable star epoch photometry which are the only publicly available \gaia time-series at the present time. The nominal scanning law will be updated in the early third \gaia data release (DR3) but the true attitude determination used in the DPAC astrometric solution will not be made available at that time\footnote{\url{https://www.cosmos.esa.int/web/gaia/earlydr3}}. Therefore, in a later paper in this series we will determine a more accurate scanning law for the period covered by DR3 by applying the methods presented in this paper to the extended variable star photometry which will become available in the full DR3.

The objective of this work in the context of the Completeness of the \gaia-verse series was to more accurately infer \gaia's true scanning law and the timings of data-taking gaps, which will be used in subsequent works to infer selection functions for the astrometric, photometric and spectroscopic data products. However, our results are also of immediate practical use. We have created a new open-source Python package \href{https://github.com/gaiaverse/scanninglaw}{\textsc{scanninglaw}} which can be used to query the times that \gaia observed a location on the sky and the probability of each of those observations resulting in a published detection. This package can download and use any of the publicly available scanning laws for the twenty two months of \gaia DR2.

\section*{Acknowledgements}
DB thanks Magdalen College for his fellowship and the Rudolf Peierls Centre for Theoretical Physics for providing office space and travel funds. AE thanks the Science and Technology Facilities Council of
the United Kingdom for financial support. This work has made use of data from the European Space Agency (ESA) mission \gaia (\url{https://www.cosmos.esa.int/gaia}), processed by the \gaia
Data Processing and Analysis Consortium (DPAC,
\url{https://www.cosmos.esa.int/web/gaia/dpac/consortium}). Funding for the DPAC
has been provided by national institutions, in particular the institutions
participating in the \gaia Multilateral Agreement.

\section*{Data availability}
The data underlying this article are publicly available from the European Space Agency's \gaia archive (\url{https://gea.esac.esa.int/archive/}). The scanning law derived in Sec. \ref{sec:scanninglaw} is publicly available on the Harvard Dataverse (\url{https://doi.org/10.7910/DVN/MYIPLH}), as are the identified gaps and detection probabilities in \gaia data-taking (\url{https://doi.org/10.7910/DVN/ST8TSM}). The authors welcome queries from those interested in using our data products in their own works.




\bibliographystyle{mnras}
\bibliography{references} 



\FloatBarrier

\appendix

\section{Matrix exponential for the MEKF}
\label{sec:matrixexponential}

The goal of this appendix is to evaluate $\exp(\delta t F)$ where $F$ is a block matrix of the form
\begin{equation}
    F = \begin{bmatrix}\omega S_3& I_3 \\ 0_3 & 0_3 \end{bmatrix},
\end{equation}
where $S_3\equiv-[\vec{n}\times]$, $\omega$ is a scalar in $\mathds{R}$ and $\vec{n}$ is a unit vector in $\mathds{R}^3$. Using the expression in Eq. 1.3 of \citet{Dieci2001} for the exponential of an upper triangular block matrix, we have
\begin{equation}
    e^{\delta t F} = \begin{bmatrix}e^{\delta t \omega S_3} & \int_0^1 \delta t e^{(1-x)\omega \delta t S_3}\mathrm{d}x \\ 0_3 & I_3\end{bmatrix}.
\end{equation}
The first diagonal term can be simplified by noting that $S_3$ is a skew-symmetric matrix with unit 2-norm and thus the Rodrigues rotation formula (e.g. see Sec. 2 of \citealp{Cardoso2010}) gives us
\begin{equation}
e^{\omega \delta t S_3} = I_3+\sin{\omega \delta t}S_3 + (1-\cos{\omega \delta t})S_3^2,
\end{equation}
where $\theta$ is a scalar in $\mathds{R}$. The off-diagonal term can be simplified by substituting the Rodrigues formula, changing variables and integrating,
\begin{align}
    \int_0^1 \delta t &e^{(1-x)\omega \delta t S_3}\mathrm{d}x = \frac{1}{\omega}\int_0^{\omega\delta t} I_3+\sin{u}S_3 + (1-\cos{u})S_3^2 \mathrm{d}u \nonumber \\
    &= \delta t I_3+\frac{1}{\omega}\left(1-\cos{\omega\delta t}\right)S_3 +\frac{1}{\omega}\left(\omega\delta t-\sin{\omega\delta t}\right)S_3^2.
\end{align}
We note that this result matches that derived by \citet{Andrle2015}.


\bsp	
\label{lastpage}
\end{document}